# Numerical Study of Oblique Detonation Initiation Assisted by Local Energy Deposition

Ziqi Jiang[a], Zongnan Chen[b], Lisong Shi[a], Zijian Zhang[c,d*], Jiaao Hao[a], Chih-yung Wen[a]

[a]Department of Aeronautical and Aviation Engineering, The Hong Kong Polytechnic University, Hong Kong S.A.R., China

[b]School of Aerospace Engineering, Beijing Institute of Technology (Zhuhai), Zhuhai 519088, China

[c]School of Aerospace Engineering, Beijing Institute of Technology, Beijing 100081, China

[d]State Key Laboratory of Environment Characteristics and Effects for Near-space, Beijing 100081, China

\* Corresponding author e-mail: zjzhang@bit.edu.cn

## Abstract

Reliable initiation of oblique detonation waves (ODWs) is crucial for the stable operation of oblique detonation engines (ODEs), especially under flight conditions of low Mach numbers and/or high altitudes. In this case, conventional initiation approaches relying solely on a fixed-angle wedge may engender risks of initiation failure, which necessitates extra initiation assistance measures. In this study, ODW initiation over a finite wedge with local energy deposition is numerically investigated to assess the thermal effects of plasma-based initiation assistance techniques. Particular emphasis is put on the effects of forms and magnitudes of energy deposition on initiation modes and flow field structures of ODWs. The results demonstrate that on-wedge initiation of ODWs fails at a low Mach number without any energy depositions. In contrast, both continuous and pulsatile local energy depositions can effectively initiate ODWs, leading to sustainable detonation on the finite wedge. As the power in continuous energy deposition increases gradually, two key detonation initiation modes emerge in sequence, namely a delayed initiation mode with combustion occurring downstream of the energy deposition zone (EDZ) and a direct initiation mode with combustion arising from the EDZ. For pulsatile energy deposition, similar delayed and direct initiation modes are found as the pulse energy increases, where the former is achieved with a local explosion spot generated through complex wave interactions downstream of the EDZ and the latter is directly generated by the EDZ. Analysis of the spatiotemporal evolution of the primary wave structures under single-pulse energy deposition reveals the minimum pulse repetition frequency required for sustainable on-wedge detonation, which is subsequently verified through multi-pulse energy deposition simulations. Nevertheless, it is found that sustainable on-wedge detonation can be achieved by pulsatile energy deposition with an average power consumption of less than 10% of that required for continuous energy deposition while maintaining a same initiation length, suggesting that the pulsatile one is an efficient way of energy deposition for initiation assistance of ODWs on finite wedges under extreme flight conditions.

**Novelty and Significance Statement:**

Initiation failure could happen when oblique detonation engines operate under wide-range flight conditions, which is imperative to address. Plasma-based ignition/combustion assistance techniques are attractive for propulsion systems but have rarely been studied in the context of oblique detonation in hypersonic premixed flow, and consequently, their applicability in initiation assistance of oblique detonation waves, as well as the relevant initiation mechanisms, remains unclear. In this study, numerical investigations of oblique detonation initiation over a finite wedge with local energy deposition are conducted to assess the thermal effects of plasma-based initiation techniques. It is the first time to have a comprehensive understanding of the initiation modes and transient flow evolutions of oblique detonation waves under the effects of both continuous and pulsatile local energy depositions with different deposition magnitudes. The critical pulse repetition frequency for sustainable on-wedge detonation initiation is proposed and verified as well. Moreover, the high efficiency of pulsatile plasma sources in oblique detonation initiation facilitation in terms of power requirement compared to the continuous ones is demonstrated. The results of this study advance the understanding of the detonation initiation phenomenon induced by plasma in hypersonic flow and also provide references to the design of wide-range oblique detonation engines.

---

# 1. Introduction

A detonation wave is a supersonic combustion wave characterized by a leading shock front tightly coupled with a thin exothermic reaction zone, resulting in an abrupt and significant change in thermodynamic properties [1]. The characteristics of detonation, such as supersonic self-sustained propagation and high efficiency in energy conversion, make it an ideal process for organizing combustion in a hypersonic propulsion system [2]. Spanning several decades from the mid-20$^{th}$ century onward, various detonation-based engines have been proposed, which can be categorized, based on how detonation waves are employed, into pulse detonation engines [3], rotating detonation engines [4,5], and oblique detonation engines (ODEs) [6,7]. Among them, the ODE has been considered one of the promising solutions for future hypersonic airbreathing propulsion due to its short combustor, simple structure, and good self-adaptability to various flow perturbations [8,9]. In a typical design of ODEs such as the one in [10], an oblique detonation wave (ODW) is formed when a high-velocity combustible mixture impinges upon a wedge, which concurrently functions as a detonation flame holder. In recent years, the feasibility of ODEs has been experimentally demonstrated in a large-scale hypersonic shock tunnel, highlighting its potential for practical applications [11,12].

Despite the significant superiorities of ODEs, critical technical challenges, such as hypersonic mixing, premature combustion prevention, and initiation and stabilization of ODWs in combustors, must be resolved for their reliable hypersonic applications [8,13,14]. Among them, detonation initiation is a prerequisite for the regular operation of an ODE, which has been extensively studied in the literature. Li et al. [15] investigated the initiation region of ODWs through numerical simulations. Then, it was visualized in the experiments conducted by Viguier et al. [16], where an oblique shock wave (OSW) and an oblique detonation surface intersect at a multi-wave point (MP) along with the formation of a primary transverse wave (PTW) and a slip line (SL). The presence of such wave structures suggests that a necessary length is required for detonation initiation, which is critical in the geometry design of an ODE combustor. Besides, the wave structures in the initiation region are closely related to the inflow parameters [17,18], especially in terms of the initiation length, which could be prolonged by 1–2 orders of magnitude under low Mach numbers or high altitudes [19,20]. However, it is necessary to broaden the flight envelope of ODEs, as they usually require other propulsion devices to accelerate to their startup speed, which is still too high to achieve [21,22]. Under wide-range flight conditions, it is of great importance to maintain the initiation length smaller than the combustor length. Otherwise, initiation failure of detonation or complete extinction of combustion may happen in the combustor, which has been extensively proven by numerical simulations of ODW initiation induced by finite wedges [23–25].

Various initiation assistance methods for ODWs have been investigated in recent studies. Fang et al. [26] utilized a wedge equipped with a blunt leading edge, and ODWs were force-initiated under the same conditions as previously failed cases with a sharp leading edge. Xiang et al. [27] numerically investigated the effectiveness and the relevant mechanism of using on-wedge blunt bumps to promote hot spot formation and ODW initiation. Han et al. [12] experimentally demonstrated the feasibility of using an on-wedge trip near the wedge tip to force ODW initiation. Qin and Zhang [28] utilized a double-cone configuration with two different cone angles to elevate the temperature and pressure within the initiation region, enabling a controllable initiation location of ODWs. These methods altered the geometric structure of the wedge/cone in essence to assist ODW initiation, which are known as passive initiation methods. In contrast, active initiation methods without altering the wedge/cone geometry, such as jet-assisted initiation methods, have also been widely studied. In the work by Li et al. [29], a transverse hot jet was introduced to the induction zone. Qin et al. [30] utilized a co-flow hot jet and demonstrated its capability of shortening the initiation length as compared to a straight wedge without the jet. Wang et al. [31] investigated the effects of jet angle and position, showing the potential of initiation control by hot jets as well. Zhang et al. [32] proposed two configurations of transverse-jet-assisted initiation in an ODE combustor for flight conditions at low Mach numbers and high altitudes, respectively. These studies have proven the feasibility of utilizing jet-assisted methods for ODW initiation within a wide flight envelope.

Recently, inspired by igniters that have been widely used in internal combustion engines, plasma-based approaches have also been utilized for ignition assistance and combustion enhancement in supersonic flows [33]. Various plasma actuators, such as plasma torches, multi-spark generators, and nanosecond pulsed discharge, have shown their capabilities in enhancing ignition, combustion, and flame stabilization [34–36]. The mechanisms of enhancing ignition and combustion differ based on the types of plasma and fuel, which can be categorized into three major aspects: thermal effect, kinetic effect, and transport effect [37]. Among them, the thermal effect plays a key role in typical plasma actuators such as focusing lasers [38,39] and nanosecond pulse discharge [40]. However, most of the aforementioned studies only focused on the plasma-based assistance methods for supersonic combustible mixtures, whereas the feasibility of such methods is rarely examined in the context of ODWs. Recently, numerical simulations have been carried out by Zhang et al. [41] to investigate the initiation characteristics of free ODWs induced by non-intrusive energy deposition. The results showed that with a moderate amount of energy deposition from the non-intrusive source, a stable free ODW can be successfully obtained. Liu et al. [42] utilized an instantaneous energy source with a diameter of 20 mm located at the end of a finite wedge, achieving successful initiations in cases with low inflow static pressure. However, the feasibility of utilizing plasma-assisted methods for ODW on-wedge initiation with practical energy deposition size, power, and forms (continuous or pulsatile) is yet to be well studied.

Therefore, this study aims to numerically investigate the thermal effect of the plasma-assisted initiation method for wedge-induced ODWs by modeling localized energy deposition on the surface of a finite wedge in simulations. The influences of energy input magnitudes and the forms (continuous or pulsatile) of local energy deposition on the initiation characteristics of ODWs are specifically examined. The remaining parts of the paper are organized as follows. The physical and mathematical models, as well as the numerical methods, are introduced in detail in Section 2. In Section 3, the initiation characteristics of ODWs with the assistance of continuous and pulsatile energy depositions are systematically examined, and various initiation modes emerging as the continuous deposition power or the pulsatile deposition energy increases are summarized and discussed. Finally, concluding remarks are summarized in Section 4.

## 2. Methodology

### 2.1 Physical model

In this study, the plasma-assisted initiation of ODWs is numerically investigated by utilizing a finite wedge in a hypersonic combustible flow, as illustrated in Fig. 1a. The physical model consists of a finite wedge with an inclined angle of 20° and a wedge length of $L_w = 60$ mm, followed by an expansion corner of 20°. In other words, the wall surface downstream of the expansion corner is parallel to the inflow direction. The inflow is a well-premixed stoichiometric $H_2$/air mixture at a Mach number of $Ma = 7$, a static temperature of $T_\infty = 300$ K, and a static pressure of $p_\infty = 50$ kPa. When the inflow interacts with the wedge, only an inert OSW without any combustion is generated on the finite wedge due to insufficient compression under the present low-Mach-number condition, which will be demonstrated later in Section 3.1. As the flow approaches the expansion corner, it undergoes a near-isentropic expansion process facilitated by expansion waves, resulting in an increase in flow velocity but a decrease in pressure, temperature, and density. This significantly suppresses combustion downstream of the expansion corner.

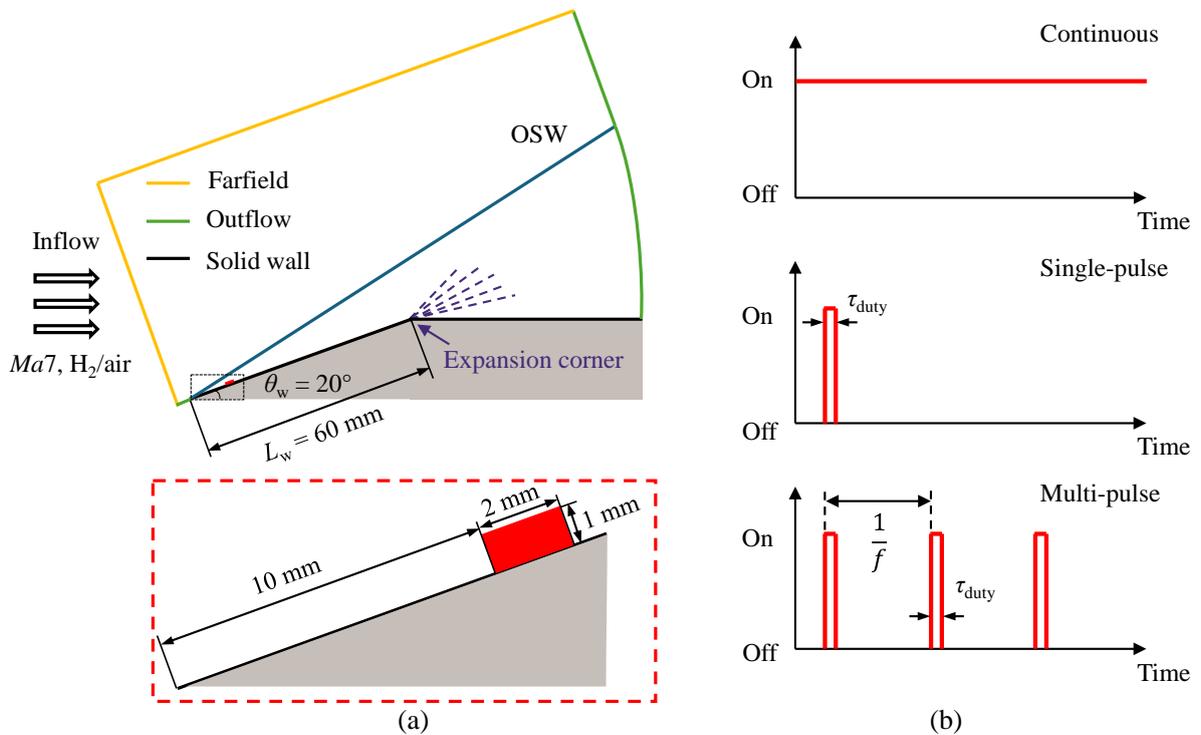

Fig. 1 Schematics of (a) a 20° finite wedge placed at a Mach 7 inflow of stoichiometric $H_2$/air mixture with a zoomed-in view of the location and geometry of the EDZ near the wedge tip and (b) different energy deposition schemes within the EDZ.

To mimic the thermal effect of plasma on ODW initiation, a small rectangular energy deposition zone (EDZ) with dimensions of 2 mm (width) × 1 mm (height) is placed 10 mm downstream from the wedge tip, where a homogenous gas heating effect is introduced, as shown in the zoomed-in view in Fig. 1a. In alignment with the two main forms of plasma actuators (i.e., continuous and pulsatile), three different time schemes of energy deposition are implemented, namely the continuous, single-pulse, and multi-pulse energy deposition schemes, as illustrated schematically in Fig. 1b. In the continuous energy deposition scheme, the EDZ remains active with constant power input. In the single-pulse energy deposition scheme, the EDZ is first activated for a period of $\tau_{duty}$ and then remains off. The multi-pulse scheme employs a periodic activation of the EDZ at a repetition frequency of $f$, forming a cycle duration of $1/f$. The activation time within each cycle is $\tau_{duty}$ as well. In this study, $\tau_{duty} = 100$ ns, which is a typical heating time for nanosecond pulse discharges. The range of $f$ for the multi-pulse deposition scheme is from 10 kHz to 50 kHz, corresponding to a duty cycle varying from 0.1% (for $f = 10$ kHz) to 0.5% (for $f = 50$ kHz).

The setting of the continuous energy deposition scheme is intended to mimic the gas-heating effect of plasma actuators such as plasma torch, arc, and other direct current discharge sources. As compared to the rapid initiation process of ODWs, the duty time of these plasma sources is typically long, which can be considered as continuously working throughout the whole initiation process [43–45]. Besides, the temperature of these plasma sources is usually as high as 10,000 K, accompanied by a high energy consumption, and hence, the thermal effect is dominant in ignition enhancement [46]. The setting of the pulsatile energy deposition in this paper corresponds to the working state of a class of high-frequency plasma actuators, for instance, high-frequency streamers [47], nanosecond discharges [48], high-frequency dielectric barrier discharges [49], and other high-frequency multi-spark generators [50]. These plasma actuators have a relatively short duty time $\tau_{duty}$ for a single pulse (at the nanosecond level) and a high frequency (at several kHz or even MHz) [51]. The duty cycle of pulsatile plasma-based sources is commonly lower than 1%. However, the total power of the pulsatile plasma-based sources is similar to or only one order of magnitude smaller than that of continuous plasma-based sources. This allows intensive deposition of energy concentrated in a brief interval, which could be beneficial to the viability of a successful ignition [52] and the reduction of induction lengths [34].

## 2.2 Mathematical models and numerical algorithms

In this study, emphasis is placed on examining the effects of both continuous and pulsatile local energy depositions for ODW initiation assistance. Following previous numerical investigations of ODW initiation assisted by other active initiation methods in literature [53–55], complex three-dimensional effects on the flow field are neglected. Therefore, the ODW flow field over the finite wedge with local energy deposition is governed by the two-dimensional (2-D) multispecies reactive Reynolds-averaged Navier-Stokes (RANS) equations, which can be written in the following conservative form.

$$\frac{\partial \mathbf{U}}{\partial t} + \frac{\partial (\mathbf{F} - \mathbf{F}_v)}{\partial x} + \frac{\partial (\mathbf{G} - \mathbf{G}_v)}{\partial y} = \mathbf{S}, \qquad (1)$$

where $\mathbf{U}$, $\mathbf{F}$, $\mathbf{F}_v$, $\mathbf{G}$, $\mathbf{G}_v$, and $\mathbf{S}$ are vectors of the conservative variables, the inviscid and viscous fluxes in the $x$-direction, the inviscid and viscous fluxes in the $y$-direction, and the source terms, respectively. They can be expressed as follows:

$$\mathbf{U} = \begin{bmatrix} \rho \\ \rho u \\ \rho v \\ \rho e \\ \rho Y_1 \\ \rho Y_2 \\ \vdots \\ \rho Y_{n_s-1} \\ \rho \tilde{v} \end{bmatrix}, \mathbf{F} = \begin{bmatrix} \rho u \\ \rho u^2 + p \\ \rho uv \\ (\rho e + p)u \\ \rho u Y_1 \\ \rho u Y_2 \\ \vdots \\ \rho u Y_{n_s-1} \\ \rho u \tilde{v} \end{bmatrix}, \mathbf{F}_v = \begin{bmatrix} 0 \\ \tau_{xx} \\ \tau_{xy} \\ u\tau_{xx} + v\tau_{xy} - q_x \\ J_{1x} \\ J_{2x} \\ \vdots \\ J_{(n_s-1)x} \\ \frac{1}{\sigma_{\tilde{v}}}(\mu \frac{\partial \tilde{v}}{\partial x} + \sqrt{\rho}\tilde{v}\frac{\partial(\sqrt{\rho}\tilde{v})}{\partial x}) \end{bmatrix}, \mathbf{G} = \begin{bmatrix} \rho v \\ \rho uv \\ \rho v^2 + p \\ (\rho e + p)v \\ \rho v Y_1 \\ \rho v Y_2 \\ \vdots \\ \rho v Y_{n_s-1} \\ \rho v \tilde{v} \end{bmatrix},$$

$$\mathbf{G}_v = \begin{bmatrix} 0 \\ \tau_{yx} \\ \tau_{yy} \\ u\tau_{yx} + v\tau_{yy} - q_y \\ J_{1y} \\ J_{2y} \\ \vdots \\ J_{(n_s-1)y} \\ \frac{1}{\sigma_{\tilde{v}}}(\mu \frac{\partial \tilde{v}}{\partial y} + \sqrt{\rho}\tilde{v}\frac{\partial(\sqrt{\rho}\tilde{v})}{\partial y}) \end{bmatrix}, \mathbf{S} = \begin{bmatrix} 0 \\ 0 \\ 0 \\ S_{EDZ} \\ \dot{\omega}_1 \\ \dot{\omega}_2 \\ \vdots \\ \dot{\omega}_{n_s-1} \\ S_{\tilde{v}} \end{bmatrix}, \quad (2)$$

where $\rho$, $p$, $u$, $v$, and $e$ are the density, pressure, velocity components in the $x$-and $y$-directions, and total energy per unit mass, respectively. The $e$ can be further expressed by

$$e = \sum_{i=1}^{n_s} Y_i h_i - \frac{p}{\rho} + \frac{1}{2}\left(u^2 + v^2\right), \tag{3}$$

In Eqs. (2) and (3), $Y_i$ is the mass fraction of species $i$ ($i$ = 1, 2, …, $n_s$), and $n_s$ is the total number of species in the mixture. The $h_i$ denotes the specific enthalpy, which is a function of the gas temperature $T$. The $p$, $\rho$, and $T$ can be correlated by using the equation of state of an ideal gas written as

$$p = \rho \sum_{i=1}^{n_s} Y_i \frac{R_u}{M_i} T, \tag{4}$$

where $R_u$ is the universal gas constant and $M_i$ is the molar mass of species $i$. The $\tau_{xx}$, $\tau_{xy}$, $\tau_{yx}$, and $\tau_{yy}$ are the shear stress tensor components and can be evaluated as follows:

$$\begin{cases} \tau_{xx} = (\mu+\mu_t) \cdot \left(\dfrac{4}{3}\dfrac{\partial u}{\partial x} - \dfrac{2}{3}\dfrac{\partial v}{\partial y}\right) \\ \tau_{yy} = (\mu+\mu_t) \cdot \left(\dfrac{4}{3}\dfrac{\partial v}{\partial y} - \dfrac{2}{3}\dfrac{\partial u}{\partial x}\right) \\ \tau_{xy} = \tau_{yx} = (\mu+\mu_t) \cdot \left(\dfrac{\partial v}{\partial x} + \dfrac{\partial u}{\partial y}\right) \end{cases} \quad (5)$$

In Eq. (5), $\mu$ is the molecular (laminar) dynamic viscosity of the gas mixture obtained from Sutherland's law, while $\mu_t$ is the turbulence dynamic viscosity that is evaluated by the modified Spalart-Allmaras (S-A) turbulence model with a compressibility correction by Edwards and Chandra [56] in a density-corrected conservative form, where $\tilde{v}$ denotes the modified turbulent kinematic viscosity and the turbulent diffusion is scaled by the coefficient $\sigma_{\tilde{v}} = 2/3$ (See Eq. (2)). Fourier's law with inter-species diffusion is adopted to calculate the heat flux with its components expressed as follows:

$$\begin{cases} q_x = -(k+k_t)\dfrac{\partial T}{\partial x} + \sum_{i=1}^{n_s} h_i J_{ix} \\ q_y = -(k+k_t)\dfrac{\partial T}{\partial y} + \sum_{i=1}^{n_s} h_i J_{iy} \end{cases}, \quad (6)$$

where $k$ and $k_t$ are the molecular (laminar) thermal conductivity and turbulent thermal conductivity, respectively. They can be evaluated as

$$k = \dfrac{\mu c_p}{Pr}, \quad k_t = \dfrac{\mu_t c_p}{Pr_t} \quad (7)$$

with $c_p$ as the specific heat capacity at constant pressure, $Pr = 0.72$ the (laminar) Prandtl number, and $Pr_t = 0.9$ the turbulent Prandtl number. In Eqs. (2) and (6), $J_{ix}$ and $J_{iy}$ are the species diffusion fluxes in the $x$- and $y$-directions, respectively, written as

$$\begin{cases} J_{ix} = -\rho(D_i + \dfrac{\mu_t}{\rho Sc_t})\dfrac{\partial Y_i}{\partial x} \\ J_{iy} = -\rho(D_i + \dfrac{\mu_t}{\rho Sc_t})\dfrac{\partial Y_i}{\partial y} \end{cases}, \quad (8)$$

where $D_i$ is the molecular diffusion coefficient of species $i$ and $Sc_t = 1.0$ the turbulent Schmidt number.

For the source terms **S** in Eq. (2), the species mass production rates $\dot{\omega}_i$ are modeled by a 9-species, 19-step elementary reaction model for $H_2$/air combustion, which was initially proposed by Jachimowski [57] and modified by Wilson & MacCormack [58]. This reaction model has been extensively utilized in numerical simulations of unsteady shock-induced combustion phenomena [59,60], scramjets [61,62], ODWs [42,55], and ODEs [6,13]. The $S_{EDZ}$ represents the energy deposition source term and is evaluated by

$$S_{\text{EDZ}} = \begin{cases} \dfrac{P_{\text{EDZ}}(t)}{A_{\text{EDZ}}}, & \text{within EDZ} \\ 0, & \text{others} \end{cases}, \tag{9}$$

where $P_{\text{EDZ}}(t)$ is the energy deposition power per unit spanwise length within the EDZ and $A_{\text{EDZ}}$ the area of the EDZ. As for the continuous energy deposition scheme,

$$P_{\text{EDZ}}(t) = P_{\text{in}}, \tag{10}$$

with $P_{\text{in}}$ as the continuous energy deposition power per unit spanwise length (in kW/cm). As for the multi-pulse energy deposition scheme with a frequency of $f$ and an activation time of $\tau_{\text{duty}}$ within each cycle, $P_{\text{EDZ}}(t)$ is evaluated by

$$P_{\text{EDZ}}(t) = \begin{cases} \dfrac{E_{\text{sp}}}{t_{\text{duty}}}, & (n-1)/f \leq t \leq (n-1)/f + t_{\text{duty}} \\ 0, & \text{others} \end{cases}, \tag{11}$$

where $E_{\text{sp}}$ is the deposition energy per unit spanwise length (in mJ/cm) of one pulse and $n$ (= 1, 2, 3, …) the cycle No. Obviously, as for the single-pulse scheme, $n \equiv 1$ in Eq. (11).

The aforementioned governing equations are numerically solved by an in-house code called Parallel Hypersonic Aerothermodynamics and Radiation Optimized Solver (PHAROS) based on the finite volume method [63,64]. The inviscid fluxes are evaluated utilizing the modified Steger-Warming scheme [65]. The second-order monotonic upstream-centered scheme for conservation laws (MUSCL) with van Leer Limiter [66] is utilized for reconstruction. The viscous fluxes are calculated using a second-order central difference scheme. A second-order implicit point relaxation method is employed for time integration [67]. This solver has been successfully used in simulations of various inert and reactive hypersonic flow problems, including nanosecond pulsed plasma [68] and detonation [69].

## 2.3 Numerical setups

The computational domain is depicted in Fig. 1a. The left and upper boundaries are imposed by the supersonic inflow conditions described in Section 2.1. The right boundary is set by the iso-gradient supersonic outflow conditions. The lower boundary is set as an isothermal wall with a wall temperature of $T_w = 300$ K. Quadrilateral grid cells with a total number of 1.26 million are utilized in this study. The maximum grid size is limited to 20 μm in the whole computational domain, and the grids cluster near the wall surface with a minimum grid height of 0.1 μm to ensure the corresponding $y^+$ value in the order of unity. Resolution studies are conducted to confirm the grid independence of the simulated flow fields. The aforementioned mesh is refined by halving the grid size in both the $x$- and $y$-directions, resulting in a total of 5.04 million grid cells with a maximum grid size of 10 μm.

The flow fields of ODWs assisted by continuous energy deposition with an input power of $P_{\text{in}}$ = 30 kW/cm, obtained using the present and refined meshes, are compared in Fig. 2. As seen, differences in flow structures are hard to identify. To further examine the effects of grid size on the numerical results, the distributions of pressure, temperature, and H$_2$ mole fraction along two different streamlines are shown in Fig. 3. It can be revealed that for both the two streamlines, the solid lines and dash lines, respectively representing the flow parameter distributions obtained using the present and refined meshes, are almost overlapped with each other, indicating that the simulated results are affected insignificantly by the difference in grid size. In other words, the present mesh is sufficient to simulate the initiation features of ODWs assisted by local energy deposition.

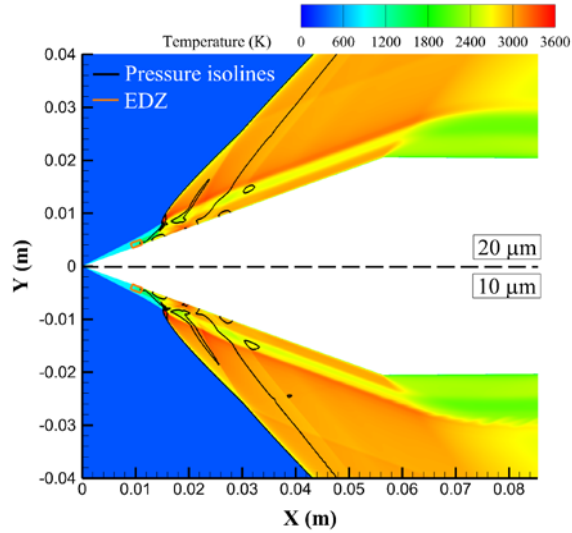

Fig. 2 Comparison of temperature contours overlaid with pressure isolines between the present mesh with a maximum grid size of 20 μm (upper) and the refined mesh with a maximum grid size of 10 μm (lower) in wedge-induced ODWs assisted with a continuous energy deposition power of Pin = 30 kW/cm.

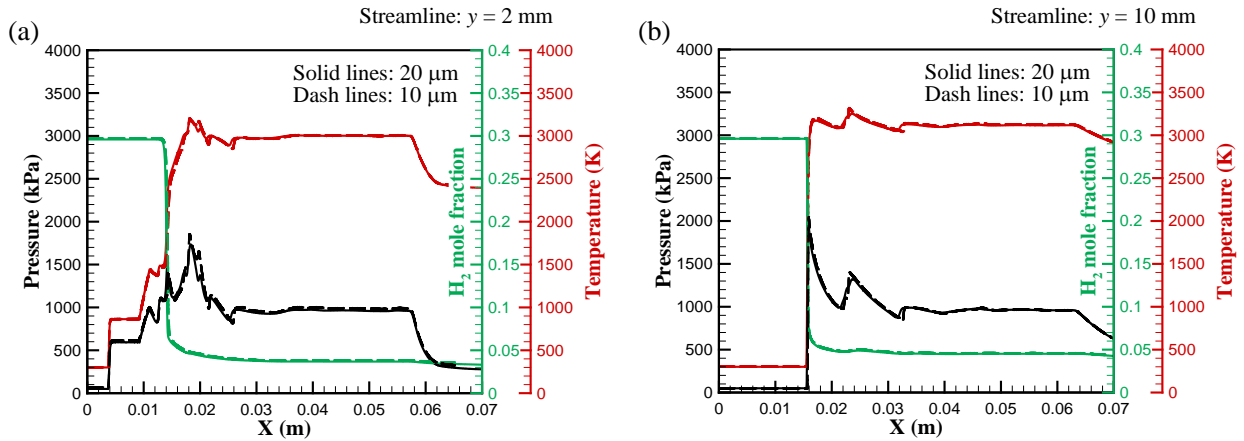

Fig. 3 Distributions of pressure, temperature, and H$_2$ mole fraction along the two streamlines marked in Fig. 2 with a continuous energy deposition power of $P_{in}$ = 30 kW/cm: (a) a streamline starting from $(x, y)$ = (0, 2) mm, and (b) a streamline starting from $(x, y)$ = (0, 10) mm.

## 3. Results and discussion

### 3.1 Continuous energy deposition

When the supersonic H$_2$/air mixture flows over the wedge, it is compressed, and an OSW is formed from the wedge tip (see the temperature contours shown in the upper part of Fig. 4a), leading to an abrupt elevation in both pressure and temperature downstream, as illustrated by the flow parameter distributions in Fig. 4b. When EDZ is not activated, the distribution of H$_2$ mole fraction downstream of the OSW remains as in the inflow (see the dashed green line in Fig. 4b), indicating absence of any combustion on the finite wedge. Downstream of the expansion corner, the flow accelerates, and both pressure and temperature drop significantly, further reducing the possibility of combustion. In other words, no combustion or initiation of detonation occurs under the present low-Mach-number inflow conditions without any energy depositions. When external energy is deposited continuously into the EDZ (see Fig. 1b), three distinct modes of the flow field (i.e., failed initiation, delayed initiation mode, and direct initiation mode) sequentially emerge as the deposition power $P_{in}$ increases, which will be discussed separately below.

### 3.1.1 Failed initiation at a low deposition power

When $P_{in} < 13$ kW/cm, as shown in the lower part of Fig. 4a by taking $P_{in} = 12$ kW/cm as an example, the continuous energy deposition within the EDZ leads to a noticeable gas heating effect, forming a thin high-temperature zone near the wall. From the distributions of flow parameters along a streamline starting from $(x, y) = (0, 2)$ mm and crossing the EDZ (see the solid lines in Fig. 4b), the gas temperature largely levels off downstream of the EDZ and forms an elevated plateau, and accordingly, the gas pressure exhibits a peak at the end of EDZ. However, the $P_{in}$ in this range is not sufficient to initiate any combustion on the finite wedge, resulting in failures of initiation. This can be concluded from the $H_2$ mole fraction distributions illustrated in Fig. 4b, where it maintains the same value as that in the inflow downstream of the EDZ.

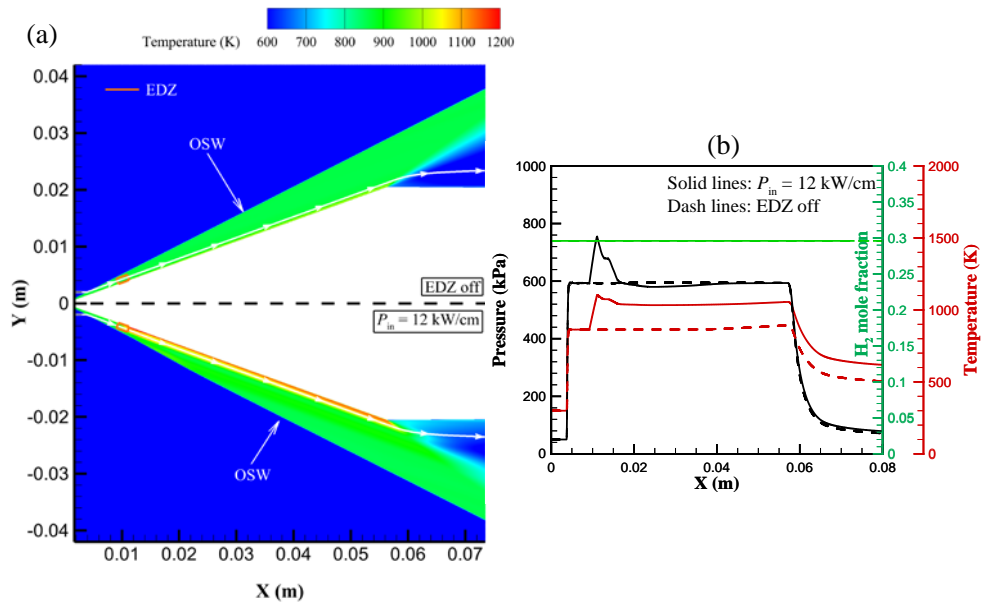

Fig. 4 (a) Temperature contours of the flow fields with ($P_{in} = 12$ kW/cm) or without a continuous energy deposition, (b) distributions of flow parameters along the streamline starting from $(x, y) = (0, 2)$ mm.

### 3.1.2 Delayed initiation mode at a moderate deposition power

Initiation of an on-wedge ODW can be achieved with $P_{in} \geq 13$ kW/cm. Here, an example of $P_{in} = 18$ kW/cm is shown in Fig. 5. The sustained energy deposition elevates the near-wall temperature sufficiently, igniting the mixture and generating a reaction front (RF) downstream of the EDZ. The RF prolongs further downstream, during which combustion within the RF is gradually accelerated with an increase in the angle of RF and a decrease in the streamwise distance to the OSW. Then, the RF intersects with the OSW, leading to significant elevations of both pressure and temperature. As a result, the RF is tightly coupled with the OSW, and an ODW is initiated. In this case, a PTW and a primary SL are also generated by the interaction between RF and OSW. Such an intersection point of OSW, ODW, PTW, and SL is known as the MP. An abrupt initiation pattern of ODW [17] is formed in this case. Although continuous energy deposition succeeds in triggering combustion and detonation before the expansion corner, these chemical reactions still take place downstream of the EDZ with a relatively long RF, and a considerable length is required for ODW initiation. Hence, an initiation mode of ODWs assisted by continuous energy deposition that exhibits the aforementioned features is classified as a delayed initiation mode in this study.

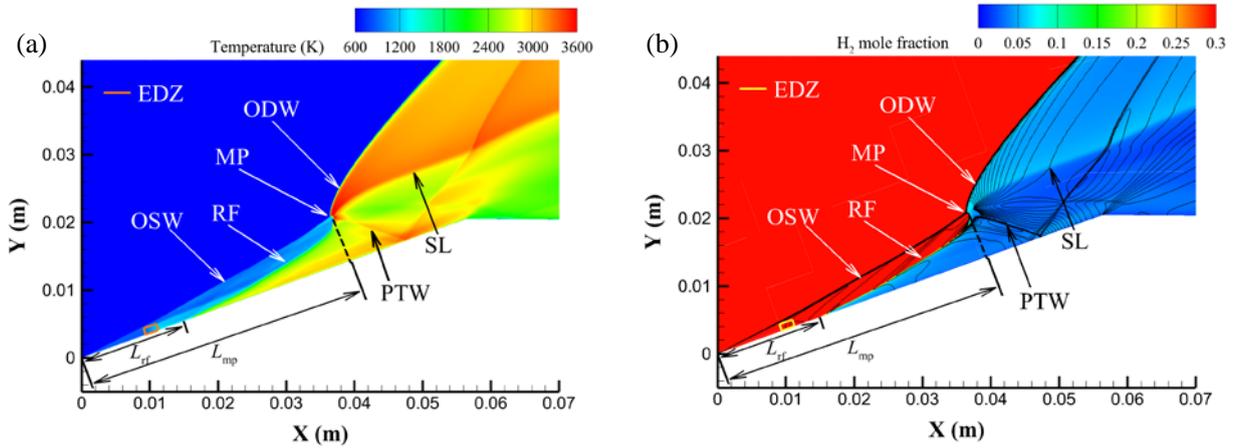

Fig. 5 Contours of (a) temperature and (b) $H_2$ mole fraction overlaid with pressure isolines of the delayed initiation mode of ODW assisted by continuous energy deposition with $P_{in}$ = 18 kW/cm.

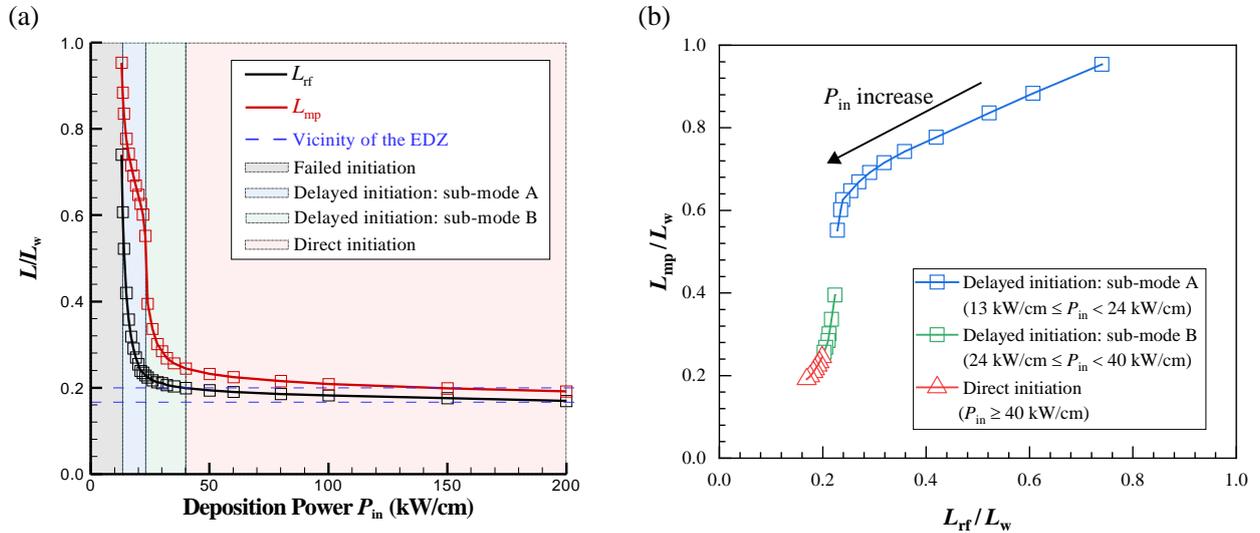

Fig. 6 (a) Variations of $L_{rf}/L_w$ and $L_{mp}/L_w$ as functions of $P_{in}$ in continuous energy deposition, and (b) the relevant variation of $L_{mp}/L_w$ as a function of $L_{rf}/L_w$.

To quantitatively evaluate the initiation features of ODWs assisted by continuous energy deposition, two characteristic lengths are defined, as shown in Fig. 5. One is the ODW initiation length $L_{mp}$, defined as the projection distance on the wedge from its leading edge to the MP. The other one is the ignition length $L_{rf}$, which is the length from the leading edge of the wedge to the root of the RF on the wedge surface. Apparently, $L_{mp}$ represents the minimum length required for ODW formation, while $L_{rf}$ denotes the minimum length required for combustion.

The variations of $L_{mp}/L_w$ and $L_{rf}/L_w$ with $P_{in}$ in continuous energy deposition are summarized in Fig. 6a. Notably, $0.167 < L/L_w < 0.2$ represents the location of the EDZ. It can be revealed that $L_{mp}$ is generally greater than $L_{rf}$ because detonation initiation arises from ignition near the wedge surface. When $P_{in} < 13$ kW/cm (i.e., the grey zone in Fig. 6a), both the dimensionless lengths $L_{rf}/L_w$ and $L_{mp}/L_w$ are greater than 1, implying that neither combustion nor detonation initiation happens before the expansion corner, namely, the failed initiation mode discussed in Section 3.1.1. As $P_{in}$ increases beyond 13 kW/cm, both $L_{rf}/L_w$ and $L_{mp}/L_w$ are smaller than 1, implying on-wedge initiation of ODWs. The $L_{rf}/L_w$ remains greater than 0.2 until $P_{in}$ increases beyond 40 kW/cm, implying that combustion happens downstream of the EDZ. In other words, the delayed initiation mode of ODWs occurs in an energy deposition power range of 13 kW/cm $\leq P_{in} <$ 40 kW/cm.

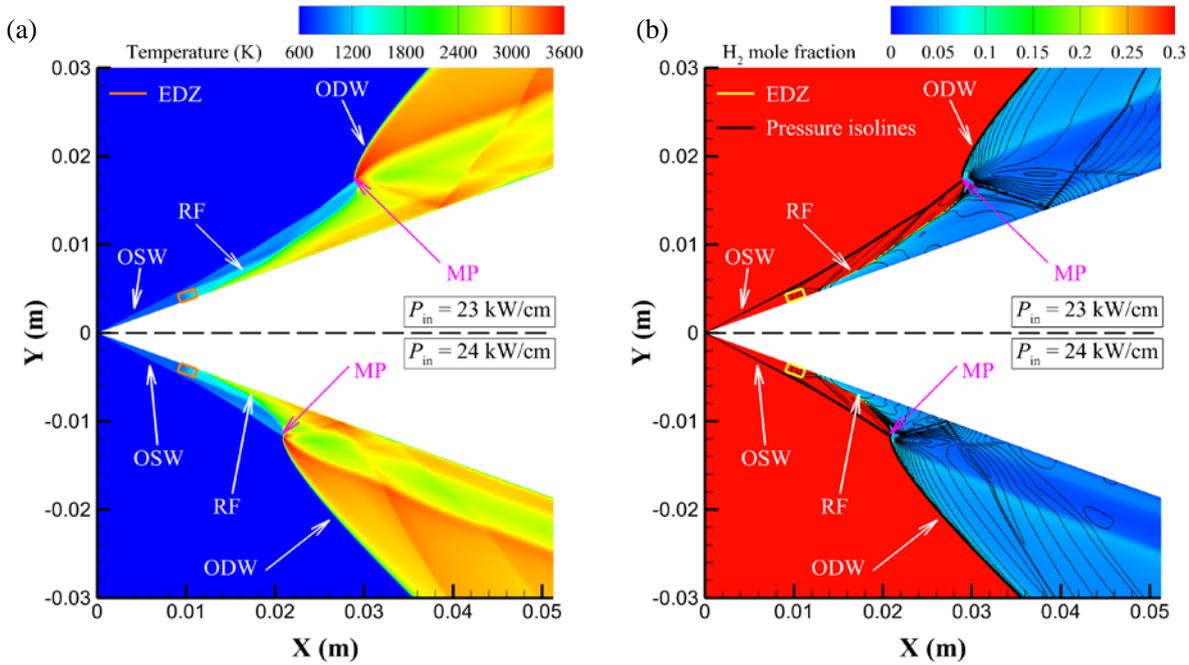

Fig. 7 Comparison of ODW flow fields assisted by continuous energy depositions with $P_{in}$ = 23 kW/cm (upper) and 24 kW/cm (lower): (a) temperature contours and (b) $H_2$ mole fraction contours with pressure isolines.

Furthermore, in the power range of delayed initiation mode, $L_{rf}/L_w$ exhibits an initial steep descent as $P_{in}$ increases, followed by a much more gradual reduction rate afterward (see Fig. 6a). In contrast, the variation of $L_{mp}/L_w$ initially manifests a more moderate decrease compared to that of $L_{rf}/L_w$, but it experiences an abrupt drop as $P_{in}$ increases beyond 24 kW/cm. After this critical point, $L_{mp}/L_w$ exhibits a similar reduction rate as $L_{rf}/L_w$. This implies two distinct sub-modes in delayed initiation of ODWs assisted by continuous energy deposition, which are referred to as delayed initiation sub-mode A (13 kW/cm $\leq P_{in} <$ 24 kW/cm, the blue zone in Fig. 6a) and sub-mode B (24 kW/cm $\leq P_{in} <$ 40 kW/cm, the green zone in Fig. 6a) in this study. The transition between these two sub-modes can be further demonstrated by the inflection point in the variation of $L_{mp}/L_w$ with $L_{rf}/L_w$ shown in Fig. 6b. As seen, the $L_{mp}/L_w$ synchronously decreases with the $L_{rf}/L_w$ at the beginning. When $L_{rf}/L_w$ is smaller than approximately 0.22 (corresponding to $P_{in} \geq$ 24 kW/cm), the $L_{mp}/L_w$ drops dramatically as $L_{rf}/L_w$ decreases. Figure 7 compares the detailed ODW flow fields of these two delayed initiation sub-modes near the transition power, i.e., $P_{in}$ = 23 kW/cm and 24 kW/cm. It can be revealed that the continuous energy deposition first generates decoupled combustion downstream of the EDZ in both cases, forming an RF at approximately the same streamwise location. However, the location where ODW forms (i.e., the MP) significantly moves upstream by 27% even if the $P_{in}$ increases slightly by only 1 kW/cm, indicating a transition from the delayed initiation sub-mode A to sub-mode B.

To unveil the main factors responsible for the dramatic advance of the MP from delayed initiation sub-mode A to sub-mode B, the flow structures near the initiation zone of these two sub-modes are further analyzed in detail. Figure 8a shows the numerical schlieren of the flow field in sub-mode A by taking the case of $P_{in}$ = 18 kW/cm as an example. It can be revealed that the energy deposition produces a significant temperature elevation in the near-wall region downstream from the EDZ, leading to the formation of an SL that is parallel to the wedge surface. The high temperature ignites the gas mixture near the wedge surface, and an RF is formed downstream of the EDZ at approximately $x$ = 0.016 m. Then, the rapid heat release within the RF results in further elevations of temperature and pressure, generating a secondary oblique shock wave (denoted as SOSW1 here) within the initiation zone upon the formation of the RF. During the process of prolonging downstream, the RF interacts with the SL, and another secondary oblique shock wave (denoted as SOSW2) is formed due to the pressure mismatch between the regions above the SL and behind the RF. Further acceleration of the chemical reaction rates within the RF is accomplished by the effects of both SOSW1 and SOSW2, and as a result, a transition from decoupled combustion to detonation is triggered downstream, generating a secondary oblique detonation wave (SODW) behind the OSW. The SODW exhibits a much larger wave angle than the RF, and it quickly intersects with the OSW, forming an MP, a PTW, and ultimately an ODW.

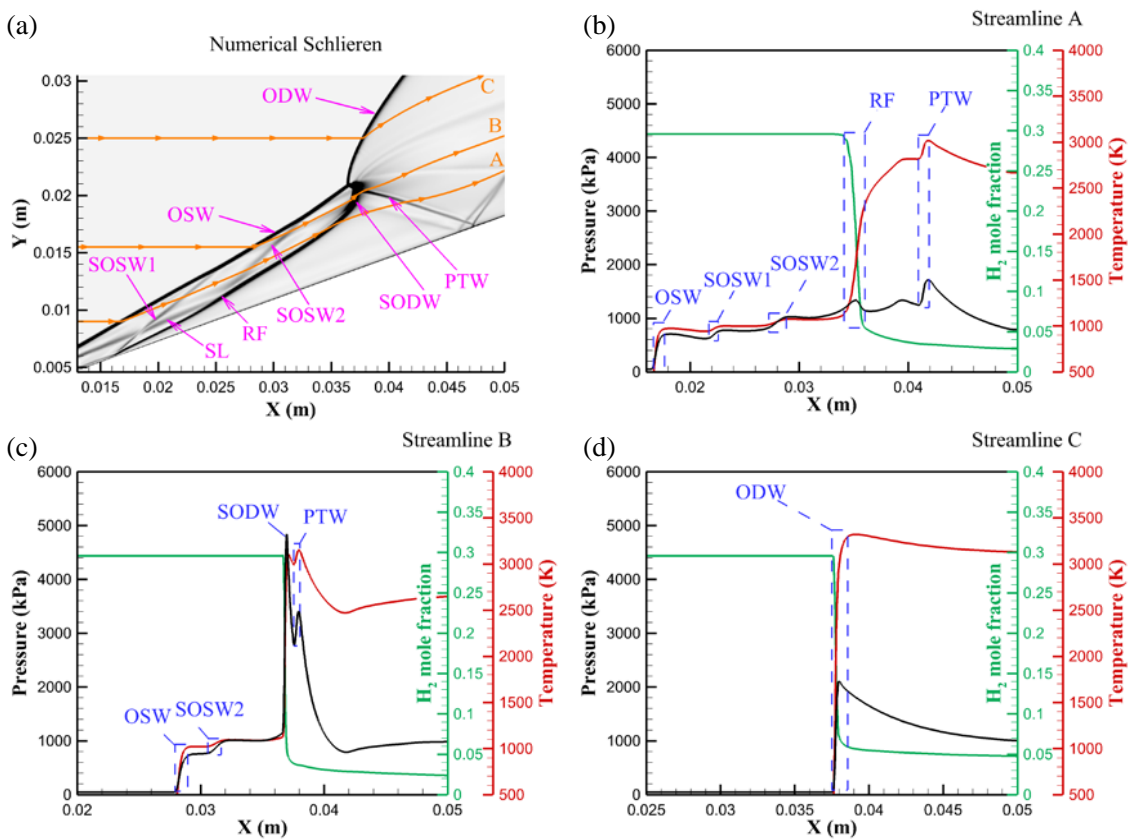

Fig. 8 ODW flow field of the delayed initiation sub-mode A at $P_{in}$ = 18 kW/cm: (a) numerical schlieren, (b–d) flow parameter distributions along three different streamlines.

The aforementioned wave structures of delayed initiation sub-mode A can be further demonstrated by flow parameter distributions along different streamlines shown in Figs. 8b-8d. Streamline A is the one travelling across the RF. Both pressure and temperature depict a three-stage increment corresponding to the existence of the OSW, the SOSW1, and the SOSW2 sequentially (see Fig. 8b). Thereafter, a rapid reduction in the $H_2$ mole fraction is noticed at the RF with a massive heat release, leading to a dramatic rise in temperature. Nonetheless, only a minimal increment in pressure is observed at this point, suggesting the absence of a shock wave and the occurrence of decoupled combustion at the RF. Further downstream, minor elevations in both pressure and temperature attributed to the PTW are observed. Streamline B crosses the SODW. Similarly, increments in pressure and temperature under the influences of the OSW and SOSW2 can be observed initially in Fig. 8c. However, different from streamline A, there is a dramatic decrease in $H_2$ mole fraction accompanied by sharp increments in both temperature and pressure for streamline B. This feature confirms the occurrence of a tight coupling between combustion and shock compression, which manifests as an SODW rather than an RF. The subsequent peaks in pressure and temperature are also caused by the PTW. Along streamline C that is away from the wedge surface, as shown in Fig. 8d, the inflow parameters remain unchanged until the successful initiation of the ODW, where the rapid consumption of the $H_2$ mole fraction is tightly coupled with substantial increases in both pressure and temperature.

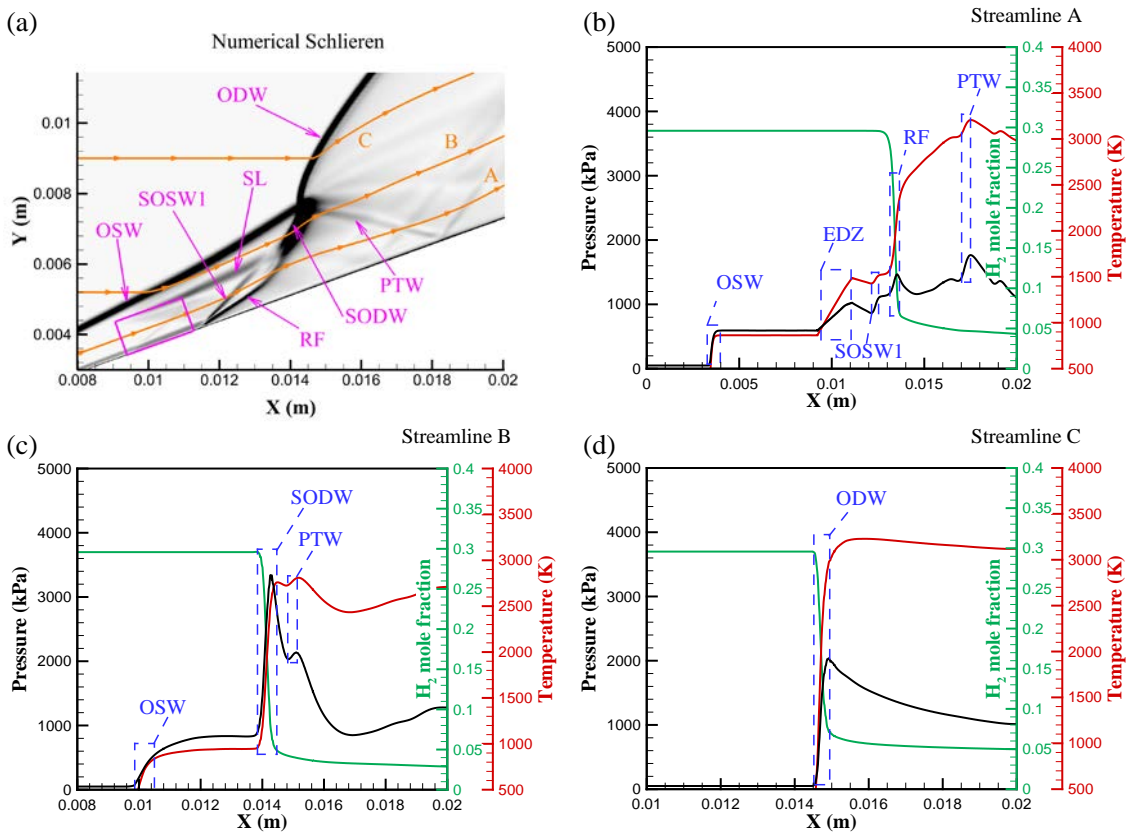

Fig. 9 ODW flow field of the delayed initiation sub-mode B at $P_{in}$ = 32 kW/cm: (a) numerical schlieren, (b–d) flow parameter distributions along three different streamlines.

The main flow structures of the delayed initiation sub-mod B are analyzed by using $P_{in}$ = 32 kW/cm as an example. As shown in the numerical schlieren in Fig. 9a, an SL is also established downstream of the EDZ owing to the significant temperature elevation by energy deposition. Similar to sub-mode A, an RF and an SOSW1 are formed downstream of the EDZ due to the ignition of the gas mixture near the wedge surface. The combustion within the RF is profoundly accelerated by the higher temperature beneath the SL, resulting in a more rapid increase in the RF angle and a more advanced intersection between the RF and the SL than those in the previous sub-mode A (see Fig. 8a). Downstream of this intersection point, the RF quickly transitions to an SODW. Subsequently, the SODW interacted with the OSW, initiating an ODW downstream and generating an MP and a PTW.

Figure 9b illustrates the distributions of flow parameters along streamline A, which traverses the RF as marked in Fig. 9a. Downstream of the OSW, noticeable increments in temperature and pressure are observed within the EDZ due to the gas heating effect, followed by the minor elevations in temperature and pressure attributed to the SOSW1. The $H_2$ mole fraction remains unchanged until reaching the RF. At the RF, $H_2$ is rapidly consumed, accompanied by a significant rise in gas temperature. Similar to that previously discussed in Fig. 8b, the increase in pressure at this point is slight, indicating decoupling of combustion from the shock waves. The other peaks in pressure and temperature observed further downstream correspond to the PTW. The distributions of flow parameters along streamline B, which goes through the SODW, are shown in Fig. 9c. Downstream of the OSW, a sharp drop in $H_2$ mole fraction and rapid increases in both temperature and pressure happen simultaneously, indicating the onset of a detonation wave, i.e., the SODW, which is distinguishably different from those observed at the RF in Fig. 9b. Similarly, the PTW also further increases the temperature and pressure downstream of the SODW. As for the parameter distributions along streamline C shown in Fig. 9d, a tight coupling between combustion and shock wave is observed, indicating the formation of the main ODW.

Through the above analysis on numerical schlieren and flow-parameter distributions along streamlines, the differences in initiation mechanism between the two delayed initiation sub-modes can be revealed. As for sub-mode A with a lower energy deposition power, the initiation of detonation results from gradual combustion acceleration caused by multiple shock compressions of OSW, SOSW1, and SOSW2. In this case, a long RF is required before its transition into detonation. Hence, the decrease in $L_{mp}$ is synchronous with $L_{rf}$ as $P_{in}$ increases for this sub-mode. As for sub-mode B, a much faster acceleration of combustion is implemented by the higher energy deposition as well as the shock compressions of OSW and SOSW1. As a result, the RF transits into an SODW (rather than the SOSW2) immediately as it goes across the SL. In other words, the RF is decoupled from SOSW2 in sub-mode A, whereas it is tightly coupled with SOSW2 to generate the SODW in sub-mode B. As the wave angle of the SODW is greater than that of the RF, a significant advance of their intersection point with the OSW, i.e., the MP, occurs in sub-mode B, and consequently, a sudden decrease in $L_{mp}$ happens as the sub-mode A transits to the sub-mode B by increasing the $P_{in}$.

*3.1.3 Direct initiation mode at a large deposition power*

As the energy deposition power $P_{in} \geq 40$ kW/cm, $L_{rf}/L_w$ or even $L_{mp}/L_w$ is smaller than 0.2, as shown by the red zone in Fig. 6a. This means that the incident RF, or even detonation, is directly formed within the EDZ. The on-wedge initiation of ODWs assisted by continuous energy deposition with such flow features is categorized as a direct initiation mode. By taking $P_{in} = 80$ kW/cm as an example, the flow field of this initiation mode is shown in Fig. 10. As seen, the extremely intensive gas heating effect caused by large-power energy deposition within the EDZ significantly increases the local gas temperature. The high temperature within the EDZ directly induces ignition of the gas mixture, accompanied by a rapid consumption of $H_2$. Meanwhile, an SODW is generated within the EDZ. Then, this SODW interacts with the OSW near the EDZ, leading to fast initiation of the main ODW in an upstream location of the wedge.

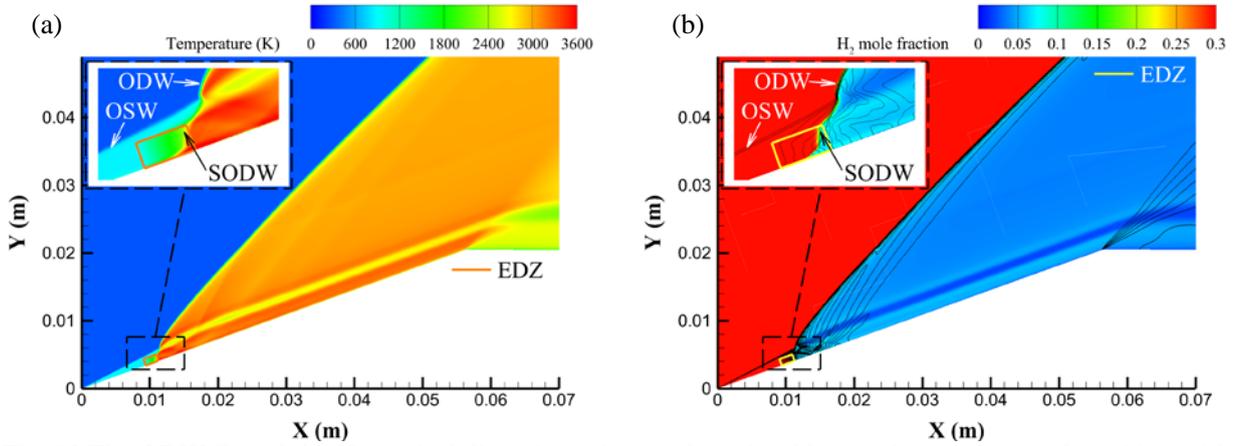

Fig. 10 The ODW flow field of a typical direct initiation mode assisted by continuous energy deposition with $P_{in}$ = 80 kW/cm: (a) temperature contours, and (b) $H_2$ mole fraction contours overlaid with pressure isolines.

Figure 11a further depicts the flow structures near the initiation zone of the direct initiation mode ($P_{in}$ = 80 kW/cm) using numerical schlieren. The SODW is clearly observed within the EDZ due to the intensive continuous energy deposition. Moreover, a distinct SL is generated at the edge of the EDZ. Following the interaction with the SL, the SODW intersects with the OSW, and consequently, the ODW is initiated. Figure 11b shows the distributions of various flow parameters along streamline A that goes across the SODW within the EDZ. After the initial elevations of pressure and temperature by the wedge-induced OSW, both pressure and temperature undergo nearly linear increases caused by the gas heating effect within the EDZ. After a concise induction period, combustion occurs with a rapid decrease in the $H_2$ mole fraction and a greater increasing rate of the temperature. Meanwhile, the pressure exhibits a distinct elevation at this location, further confirming the formation of the SODW within the EDZ. Streamline B travels across the ODW, along which the flow-parameter distributions are shown in Fig. 11c. Both pressure and temperature exhibit a significant increase, while the $H_2$ mole fraction drops, indicating a tight coupling between combustion and the shock wave, i.e., the ODW.

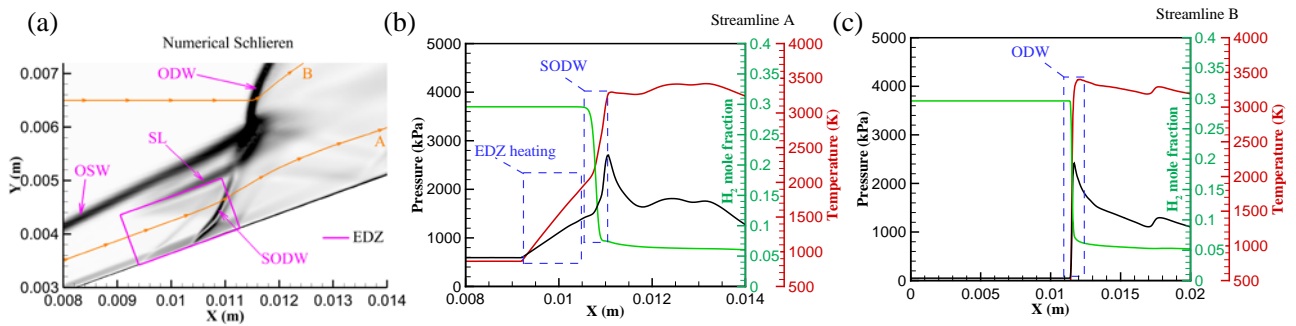

Fig. 11 ODW flow field of the direct initiation mode at $P_{in}$ = 80 kW/cm: (a) numerical schlieren, (b–c) flow parameter distributions along two different streamlines.

## 3.2. Single-pulse energy deposition

In order to have a better understanding of ODW initiation assisted by pulsatile energy deposition, the transient evolution of the key flow structures under the single-pulse scheme (see the lower part in Fig. 1c) is investigated first in this section. Similar to the continuous energy deposition, three modes emerge sequentially as the single-pulse deposition energy $E_{sp}$ increases, namely, the failed initiation, the delayed explosion-spot mode, and the direct explosion-spot mode, which will be discussed in detail below.

*3.2.1. Failed initiation at a low pulse energy*

At $E_{sp} < 15$ mJ/cm, no combustion could happen throughout the whole computational domain. When the $E_{sp}$ exceeds 15 mJ/cm, decoupled combustion occurs on the finite wedge, but it still fails to initiate an ODW before reaching the expansion corner. Taking $E_{sp} = 15$ mJ/cm as an example, the evolution of the flow field is shown in Fig. 12 by using numerical schlieren. Here, the commencement of the single-pulse energy deposition is defined as $t = 0$ (see the lower part in Fig. 1c). As depicted in Fig. 12a, no combustion happens immediately after the single pulse of energy deposited into the flow field. In this stage, compression waves (CWs) are generated by the residual heat from the energy deposition and propagate outwards. At approximately $t = 15$ μs (see Fig. 12b), ignition of the gas mixture is achieved downstream of the EDZ. A small flaming area is generated in the near-wall region following the trail of the residual heat. The flaming area keeps expanding as it propagates downstream, during which an RF is entrained out at the upstream edge of the flaming area due to the diffusion effect. Meanwhile, the rapid energy release of combustion leads to the occurrence of an SOSW. As time goes further, the flaming area quickly expands along with the prolongations of the RF and the SOSW (see Fig. 12c). Then, the RF continuously propagates downstream (see Fig. 12d), but it fails to initiate any on-wedge detonation. Finally, as both the RF and the SOSW propagate downstream of the expansion corner, there is only OSW left over the wedge, as shown in Fig. 12e. Similar flow features are observed when $E_{sp} < 22$ mJ/cm.

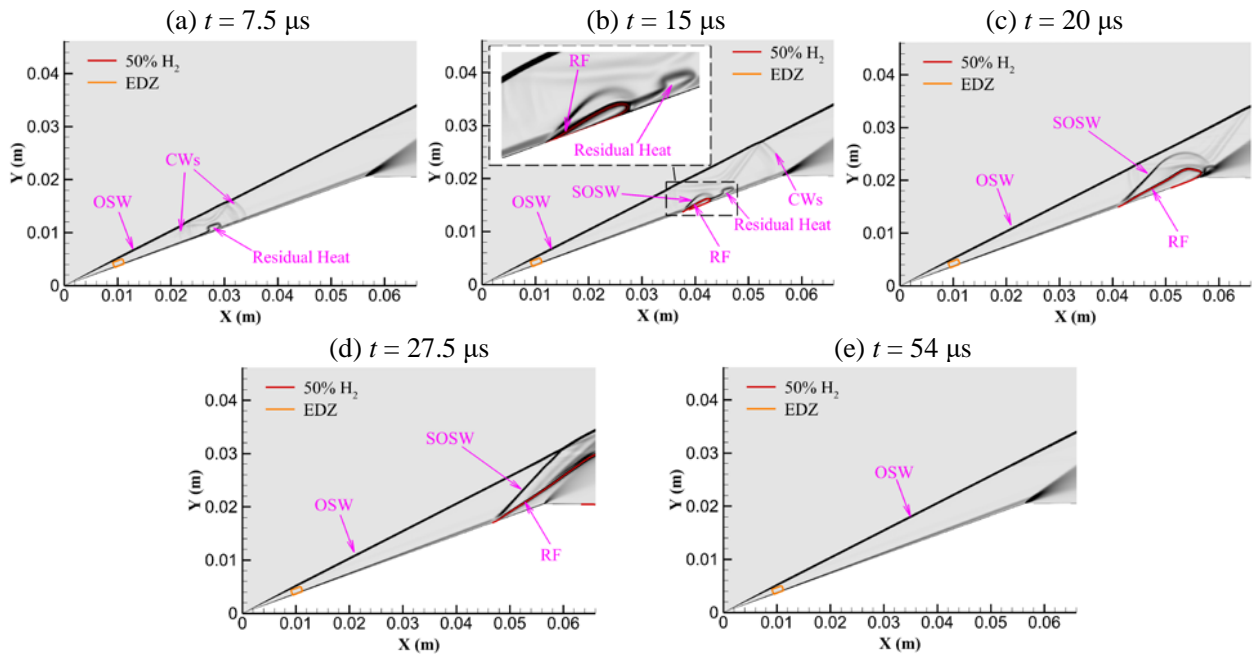

Fig. 12 Evolution of the flow field, illustrated using numerical schlieren, with a single-pulse energy deposition of $E_{sp} = 15$ mJ/cm.

*3.2.2. Delayed explosion-spot mode at a moderate pulse energy*

As $E_{sp} \geq 22$ mJ/cm, on-wedge initiation can be realized. The evolution of the flow field at $E_{sp} = 25$ mJ/cm is shown in Fig. 13 as an example. After a single pulse of energy is deposited into the flow, combustion happens downstream of the EDZ, as shown in Fig. 13a. A flaming area is formed and then expands and propagates downstream. Similar to the case previously shown in Fig. 12, an RF is entrained out at the upstream edge of the flaming area in this stage. At $t = 11.5$ μs (see Fig. 13b), the flaming area further expands, generating shock waves in each expanding direction. Among these shock waves, an SOSW is generated at the root of the RF and intersects with the OSW, reflecting an entropy wave (denoted as EW1 in Fig. 13) propagating towards the flaming area. A radially propagating shock wave interacts with the OSW, bending the segment of the OSW located above the flaming area into a curved shock wave (CSW). The last shock wave (denoted as SW here) is generated by the expanding flaming area located far downstream. Upon its interaction with the OSW, another reflected entropy wave (denoted as the EW2) is formed. Figure. 14a shows the flow-parameter variations along the streamline that goes through the flaming area, as marked by the green line with arrows in Fig. 13b. Both pressure and temperature increase after crossing the OSW. Upstream of the flaming area, the local temperature increases under the effects of the EW1, but ignition of the mixture does not occur. At the edge of the flaming area, the $H_2$ mole fraction is reduced with a significant rise in temperature, but the pressure only exhibits a little change, implying the occurrence of decoupled combustion rather than detonation at this time instant.

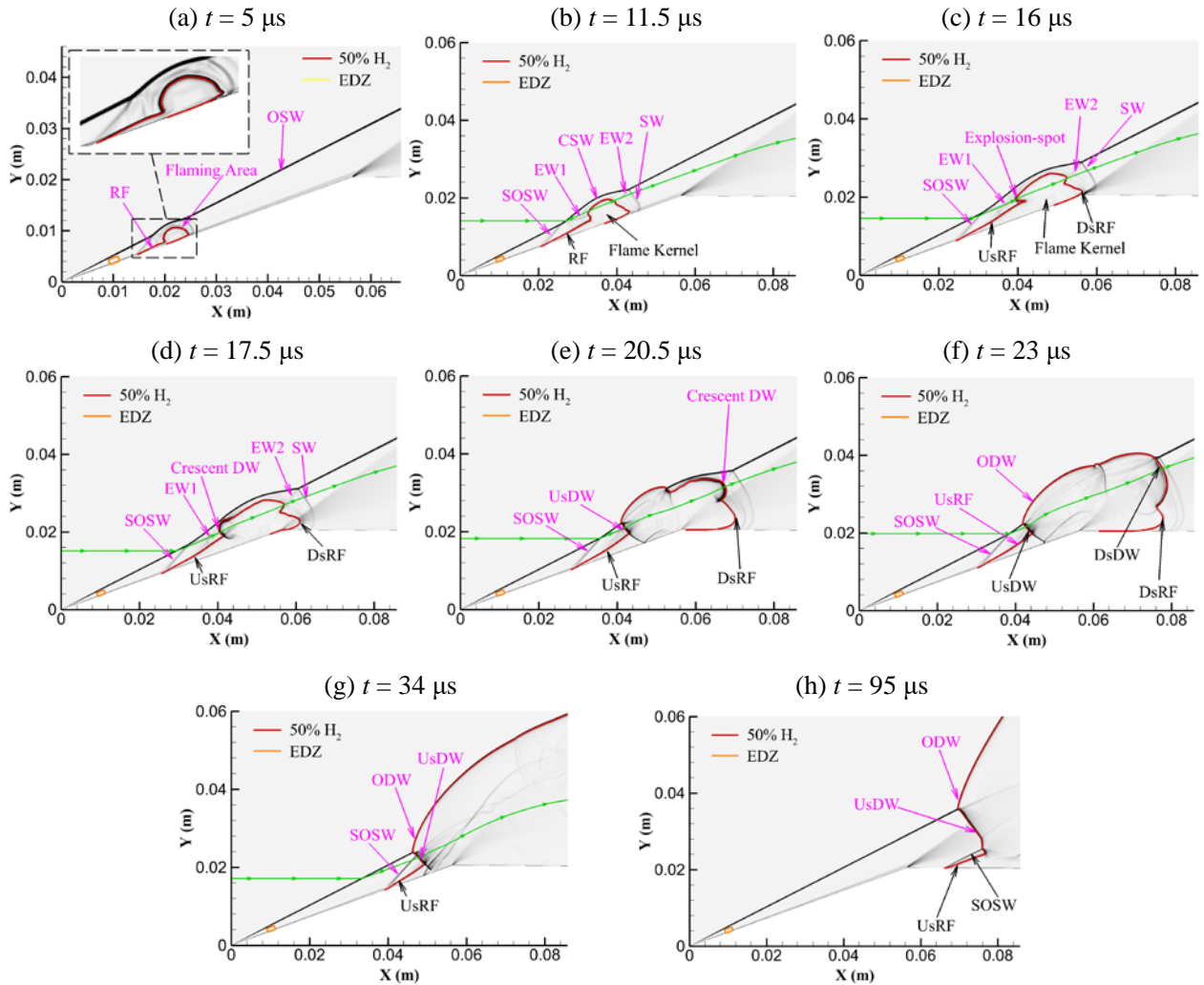

Fig. 13 Evolution of the flow field, illustrated using numerical schlieren, with a single-pulse energy deposition of $E_{sp} = 25$ mJ/cm.

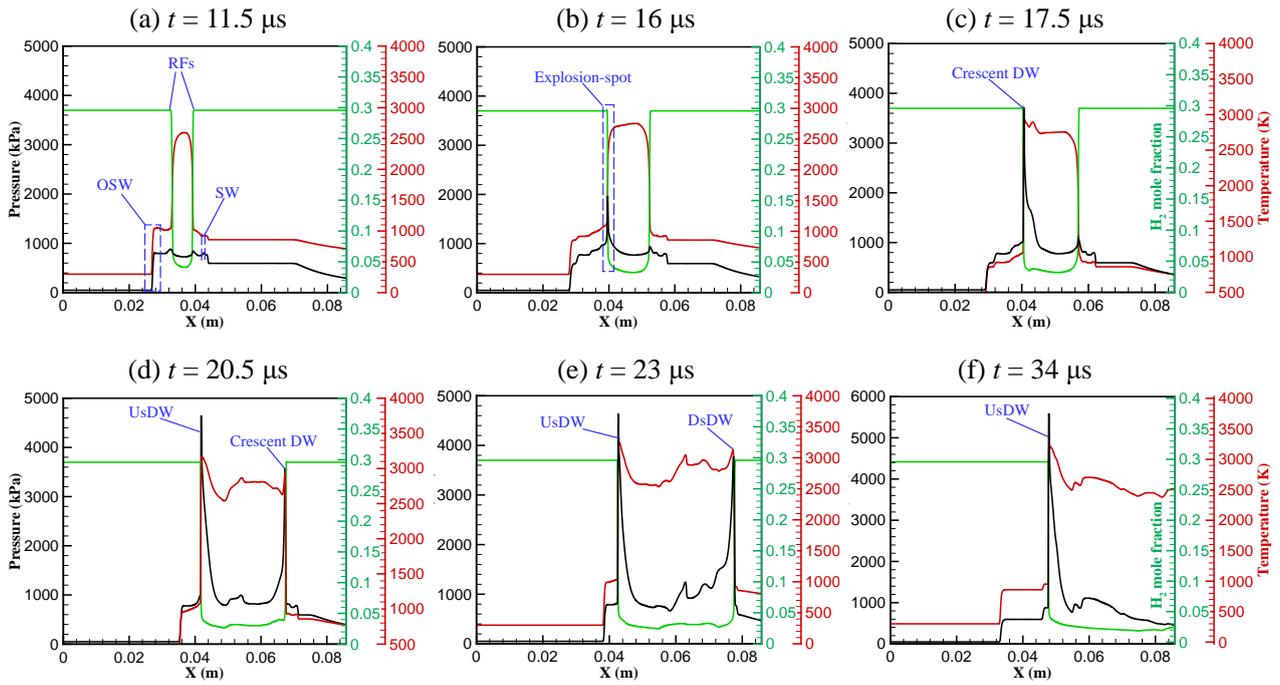

Fig. 14 Flow-parameter distributions along streamlines marked in Fig. 13 in single-pulse energy deposition with $E_{sp} = 25$ mJ/cm.

The temperature elevation caused by the EW1 accelerates the induction process of combustion upstream of the flaming area. At $t = 16$ μs (see Fig. 13c), an explosion spot is induced at the intersection point between the EW1 and the flaming area. For clarity, the RFs located upstream and downstream of the flaming area are denoted as the upstream reaction front (UsRF) and the downstream reaction front (DsRF) in this figure, respectively. The streamline distributions of flow parameters are shown in Fig. 14b. It can be observed that the pressure at the upstream edge of the flaming area begins to increase sharply, corresponding to the formation of the explosion spot. At this point, chemical reactions are accelerated by the action of the EW1, leading to a steeper rising rate of temperature. Further at $t = 17.5$ μs (see Fig. 13d), a crescent detonation wave (crescent DW) emerges at the explosion spot. In the streamline distributions of flow parameters shown in Fig. 14c, the decrease in $H_2$ mole fraction is accompanied by dramatic increases in pressure and temperature, demonstrating the onset of this crescent DW.

At $t = 20.5$ μs (see Fig. 13e), the previous crescent DW interacts with the OSW, forming an ODW and MPs at the intersection points. The crescent DW evolves into an upstream detonation wave (UsDW) connecting to the upstream MP. Meanwhile, another crescent DW is initiated at the downstream edge of the flaming area, which is induced by the action of the EW2 similarly. Both the UsDW and the downstream crescent DW can be clearly identified from Fig. 14d through the tight coupling between the significant pressure peaks and the rapid consumption of $H_2$. At $t = 23$ μs (see Fig. 13f), the downstream crescent DW also merges with the OSW, becoming a downstream detonation wave (DsDW). By comparing the streamline parameters shown in Fig. 14d and 14e, it can be revealed that both the UsDW and the DsDW propagate downstream, but the downstream-propagation velocity of the DsDW is significantly faster than that of the UsDW. After the DsDW propagating out of the computational domain (see Fig. 13g and Fig. 14f), only the upstream wave structures, including the UsRF, UsDW, SOSW, and ODW, are reserved, but they still slowly propagate downstream. Later at $t = 95$ μs (see Fig. 13h), all upstream wave structures propagate downstream of the expansion corner. As time advances further, only the OSW will be left on the wedge, and the flow field returns to a state of failure of on-wedge initiation.

Figure 15a shows the spatiotemporal evolutions of the locations of the key combustion and detonation waves, i.e., the UsRF, the DsRF, the UsDW, and the DsDW, at $E_{sp} = 25$ mJ/cm. Here, the projection locations of these wave structures on the wedge are employed for clarification since the flow behind the OSW is parallel to the wedge surface. As seen, both the UsRF and the DsRF are rapidly induced at approximately 0.015 m downstream of the leading edge of the wedge tip within $t = 5$ μs. Then, the first detonation wave, i.e., the UsDW, is initiated from the explosion spot at approximately 0.035 m downstream from the leading edge of the wedge. After that, the DsDW is initiated at a further downstream location of approximately 0.05 m. Notably, the EDZ is located approximately 0.01 m downstream from the wedge tip. The feature of detonation initiation occurring downstream in this mode is further demonstrated.

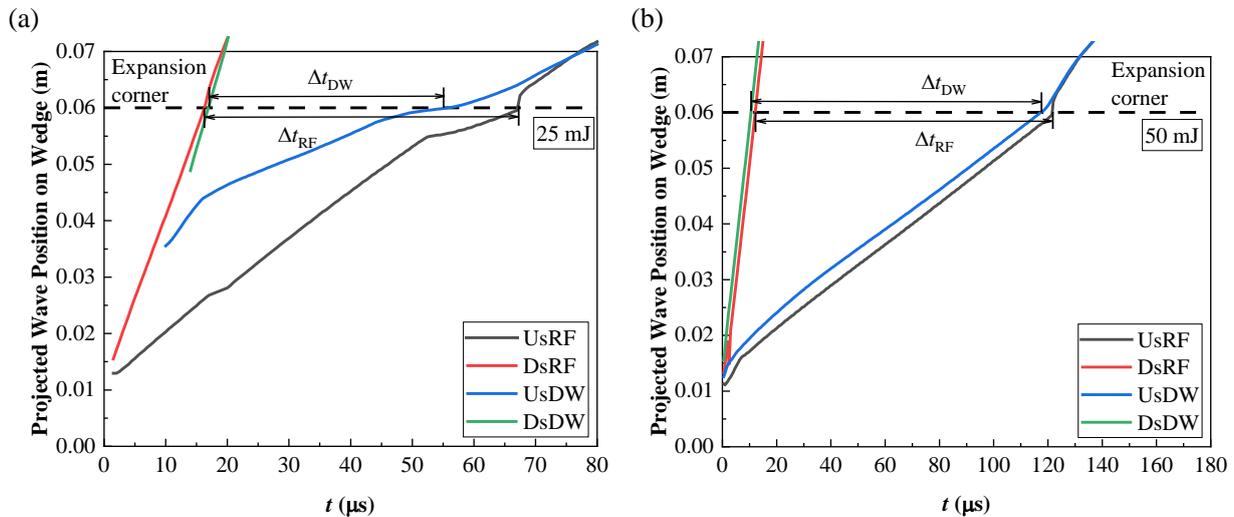

Fig. 15 Spatiotemporal evolutions of the locations of the key wave structures: (a) the delayed explosion-spot mode ($E_{sp} = 25$ mJ), and (b) the direct explosion-spot mode ($E_{sp} = 50$ mJ).

Based on the aforementioned analysis, it can be concluded that the moderate pulse energy deposition within the EDZ does not induce the initiation of detonation directly in this mode. Instead, detonation initiation is realized through the formation of explosion spots downstream of the EDZ, while these explosion spots are generated through complex wave interactions. Specifically, the explosion spots are generated through chemical reaction acceleration by the elevated temperature of the reflecting EWs from the OSW. Thus, this on-wedge initiation mode is classified as the delayed explosion-spot mode. By increasing the pulse deposition energy $E_{sp}$, it is revealed that this mode remains until $E_{sp}$ exceeds 27 mJ/cm, that is, its occurrence in the range of 22 mJ/cm $\leq E_{sp} <$ 27 mJ/cm.

*3.2.3 Direct explosion-spot mode at a large pulse energy*

The transient evolution of the ODW flow field assisted by single-pulse energy deposition of the direct explosion-spot mode ($E_{sp} \geq 27$ mJ/cm) is investigated in this section. By taking the case of $E_{sp} = 50$ mJ/cm as an example, Fig. 16 shows the instantaneous numerical schlieren photographs overlaid with half $H_2$ mole fraction isolines at different time instants. After the pulse energy is deposited within the EDZ, fast ignition of the gas mixture is induced (see Fig. 16a at $t = 1$ μs). A flaming area is formed and then gradually propagates downstream. As shown in Fig. 17a, only moderate increases in pressure are observed within the flaming area featured by rapid consumption of $H_2$, implying decoupled combustion at this moment. Later at $t = 3$ μs (see Fig. 16b), the upper edge of the flaming area interacted with the OSW, forming a curved detonation wave, which will transition to the ODW later. The upstream and downstream edges of the flaming area are coupled with shock waves, leading to formations of the UsDW and the DsDW. This feature can also be demonstrated by the distributions of flow parameters along the streamline shown in Fig. 17b, where the reduction in $H_2$ mole fraction is accompanied by significant increases in pressure and temperature. At $t$ = 11.5 µs, as shown in Fig. 16c, the flaming area is largely expanded by the UsDW and the DsDW. Relative to the local gas flow, the UsDW propagates upstream, while the DsDW propagates downstream. Concurrently, both the UsDW and the DsDW are maintained to be detonation waves throughout their propagation. The occurrence of tightly coupled detonations can be confirmed by the flow parameter distributions along the streamline shown in Fig. 17c as well.

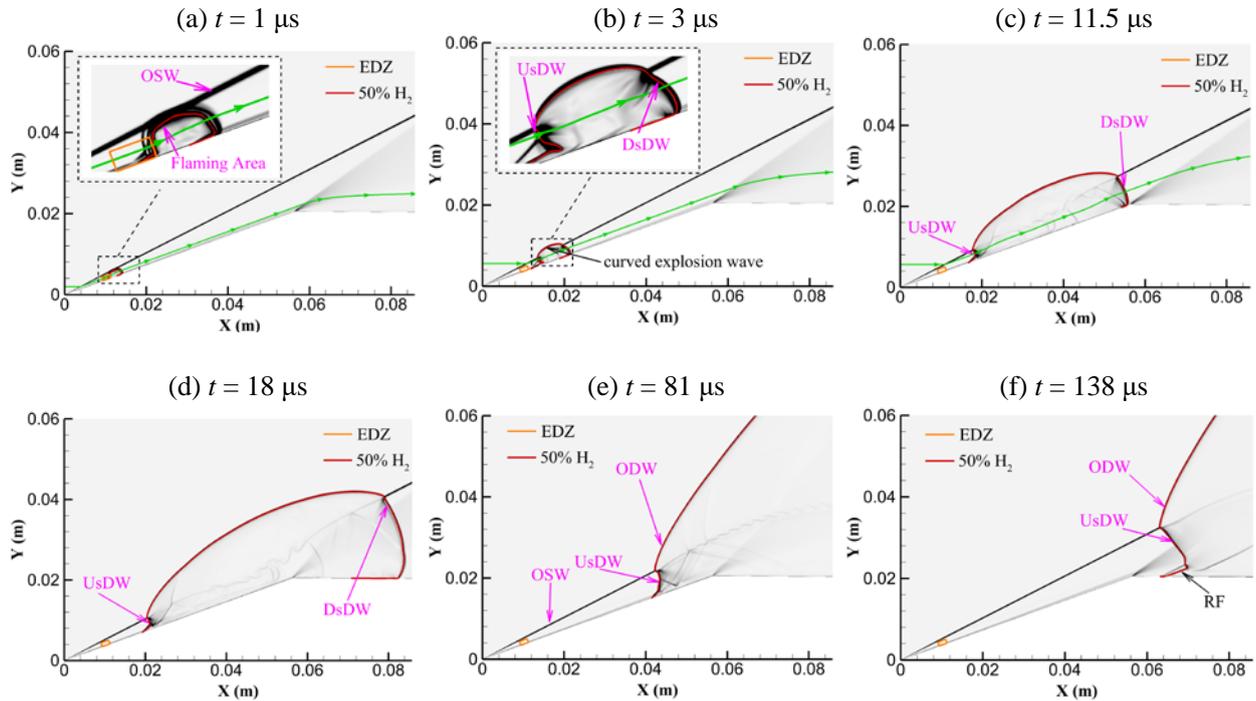

Fig. 16 Evolution of the flow field, illustrated using numerical schlieren, with a single-pulse energy deposition of $E_{sp}$ = 50 mJ/cm.

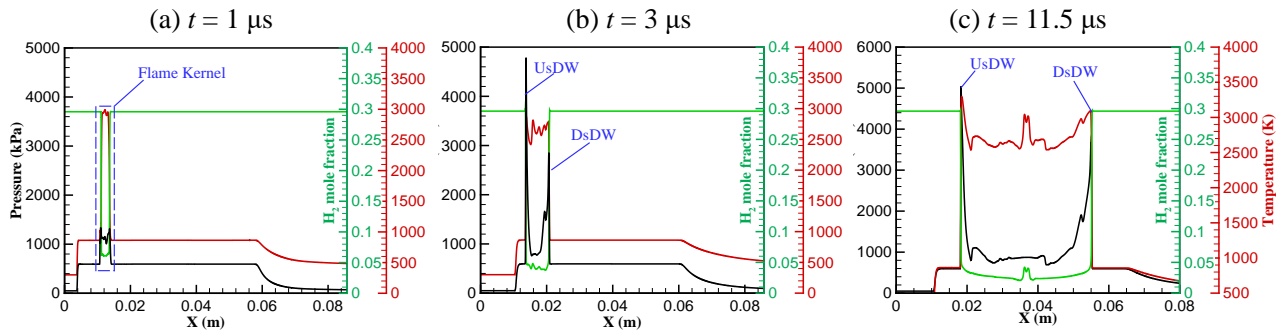

Fig. 17 Flow-parameter distributions along streamlines marked in Fig. 16 in single-pulse energy deposition with $E_{sp}$ = 50 mJ/cm.

As the downstream-propagation velocity of the DsDW is significantly greater than that of the UsDW in the wedge-fixed reference frame, the DsDW first propagates downstream of the expansion corner (see Fig. 16d at $t$ = 18 µs). At this instant, the intensities of compression and heat release of the DsDW are significantly reduced by the expansion waves. However, tight coupling between the shock wave and combustion in the DsDW is still maintained. After the complete exiting of the DsDW from the computational domain, only the UsDW remains confined to the wedge surface ($t = 81\mu s$, Fig. 16e). As time goes further, the UsDW propagates downstream of the expansion corner as well (see Fig. 16f at $t = 138\ \mu s$). After the UsDW moves out of the computational domain, only the OSW is left on the wedge eventually.

The spatiotemporal evolutions of the locations of RFs and DWs at $E_{sp} = 50$ mJ/cm are further analyzed in Fig. 15b. DWs and RFs initiate nearly synchronously, with an onset time less than $t = 3\ \mu s$, in both upstream and downstream locations. This means that the initiation of DWs is an extremely rapid process in the present mode as compared to that in the delayed explosion-spot mode. After the initiation of DWs, the UsDW and the DsDW exhibited similar trends with their corresponding RFs because of the tightly coupled combustion. The difference between the locations of the UsRF and the UsDW at the same moment is contributed by the viscous effect of the wall that entrains and prolongs the UsRF, while the most upstream point of the UsRF is used to identify its location.

The above analysis reveals that combustion and subsequently detonation initiation are induced by the significant amount of energy deposition into the flow field within the EDZ in this mode, and the intermediate process of EW interaction occurring in the delayed explosion-spot mode is not observed. Hence, this mode at $E_{sp} \geq 27$ mJ/cm is classified as the direct explosion-spot mode. It is worth noting that, as detonation initiation is induced by complex wave interactions occurring downstream of the EDZ in the delayed explosion-spot mode, the initiation location controlled by the placement of the EDZ is difficult. Instead, the reliable initiation of detonation near the EDZ in the direct explosion-spot mode makes initiation control easier, which is more favorable in practical applications.

## 3.3 Multi-pulse energy deposition

To initiate sustainable on-wedge ODWs, multi-pulse energy deposition with an appropriate pulse repetition frequency $f$ must be employed. It is expected that the initiation of ODW assisted by pulsatile energy deposition exhibits spatiotemporal self-similarity among different pulses, which means that the evolutions of key flow features such as the UsRF (or the UsDW) and the DsRF (or the DsDW) generated by the second pulse or afterwards will be the same as those produced by the first pulse. Therefore, the main wave structures and their positions in multi-pulse energy deposition can be deduced through the previous analysis of the transient evolutions of key flow structures under single-pulse energy deposition in Section 3.2.

Utilizing the direct explosion-spot mode as an example (see Fig. 18), if the UsDW by the $n^{th}$ pulse can be overtaken by (or collide with) the DsDW from the $(n+1)^{th}$ pulse before the expansion corner, and if both the distance and the time taken by the collision are short enough, it can be expected that a sustainable on-wedge ODW with a continuous detonation surface can occur. With the spatiotemporal evolutions of key wave structures shown in Fig. 15, it is easy to reveal that the minimum pulse repetition frequency $f_{cri,DW}$ for collision of the UsDW by the $n^{th}$ pulse and the DsDW by the $(n+1)^{th}$ pulse before the end of the wedge equals the inverse of the arriving time interval between them $\Delta t_{DW}$, as marked in Fig. 15, i.e., $f_{cri,DW} = 1/\Delta t_{DW}$. Similarly, sustainable on-wedge combustion can be achieved with the collision of the $n^{th}$ UsRF and the $(n+1)^{th}$ DsRF before the expansion corner. Accordingly, the minimum pulse repetition frequency for such an on-wedge collision of the RFs is given by $f_{cri,RF} = 1/\Delta t_{RF}$, where $\Delta t_{RF}$ is the arriving interval between the UsRF and the DsRF at the end of the wedge (also see Fig. 15).

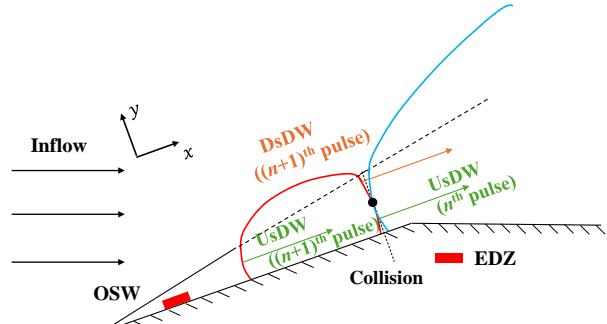

Fig. 18 A schematic of the UsDW generated by the $n^{th}$ pulse chased up by the DsDW generated by the $(n+1)^{th}$ pulse on the finite wedge under multi-pulse energy deposition.

Figure 19 shows the variations of $f_{cri,DW}$ and $f_{cri,RF}$ at different $E_{sp}$. At 22 mJ/cm $\leq E_{sp} <$ 27 mJ/cm, the DWs are initiated by the pulsatile energy deposition in the delayed explosion-spot mode. In this stage, the $f_{cri,DW}$ drops rapidly as the $E_{sp}$ increases. However, as $E_{sp} \geq$ 27 mJ/cm where the DWs are initiated in the direct explosion-spot mode, the $f_{cri,DW}$ exhibits a far slower decreasing rate and converges to a value of 8.5 kHz as the $E_{sp}$ increases to an extremely large value. This could be attributed to the nearly consistent initiation locations of the DWs in the direct explosion-spot mode with different $E_{sp}$ as well as the self-sustained propagation of detonation with nearly constant speeds with respect to local gas flow. Similarly, the $f_{cri,RF}$ first exhibits a slow decrease followed by a rapid decrease as the $E_{sp}$ increases, and then, it asymptotically approaches a fixed value at large $E_{sp}$. Furthermore, at low $E_{sp}$ in the delayed explosion-spot mode, the $f_{cri,RF}$ is significantly smaller than the $f_{cri,DW}$ due to the delayed initiation of the DWs compared to the RFs. As for the direct explosion-spot mode, the $f_{cri,RF}$ is close to the $f_{cri,DW}$ due to nearly consistent spatiotemporal evolutions between RFs and DWs in this mode.

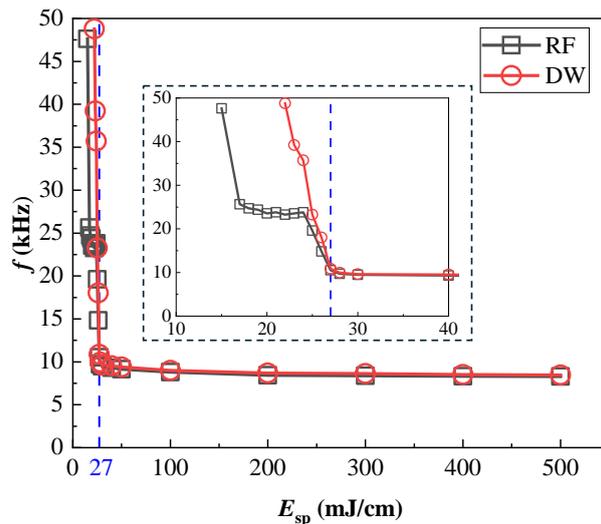

Fig. 19 Variations of the minimum pulse repetition frequencies required for on-wedge collisions of the RFs, i.e., $f_{cri,RF}$, and of the DWs, i.e., $f_{cri,DW}$, as functions of $E_{sp}$.

*3.3.1 Direct explosion-spot mode by multi-pulse energy deposition*

Taking $E_{sp}$ = 50 mJ/cm as an example, Fig. 20 shows the transient evolution of the flow field after the $(n+1)^{th}$ pulse in the direct explosion-spot mode with a pulse repetition frequency of $f$ = 10 kHz, which is only slightly higher than the $f_{cri,DW}$ = 9.4 kHz for on-wedge collision of the DWs at this $E_{sp}$. At $t = n/f + 1$ μs (see Fig. 20a), the flaming area is quickly generated within the EDZ after a large amount of energy deposition and gradually propagates downstream. At this point, the UsDW generated by the $n^{th}$ pulse remains on the wedge due to its slow propagation speed with respect to the wedge. At $t = n/f + 9$ μs (see Fig. 20b), both the UsDW and the DsDW are initiated by the intensive gas heating effect from the $(n+1)^{th}$ pulse, and the DsDW fast propagates downstream and approaches the UsDW of the $n^{th}$ pulse. At $t = n/f + 11.5$ μs (see Fig. 20c), the UsDW of the $n^{th}$ pulse is caught up by the DsDW from the $(n+1)^{th}$ pulse slightly upstream of the expansion corner. This verifies the aforementioned estimation method of $f_{cri,DW}$. Such collisions of the DWs generate a curved DW afterward, as shown in Fig. 20d. This curved DW quickly expands its length and becomes straight as it propagates downstream. Later, the ODW generated by the $(n+1)^{th}$ pulse becomes the dominant detonation surface on the wedge and keeps propagating downstream until the next pulse is generated in the EDZ.

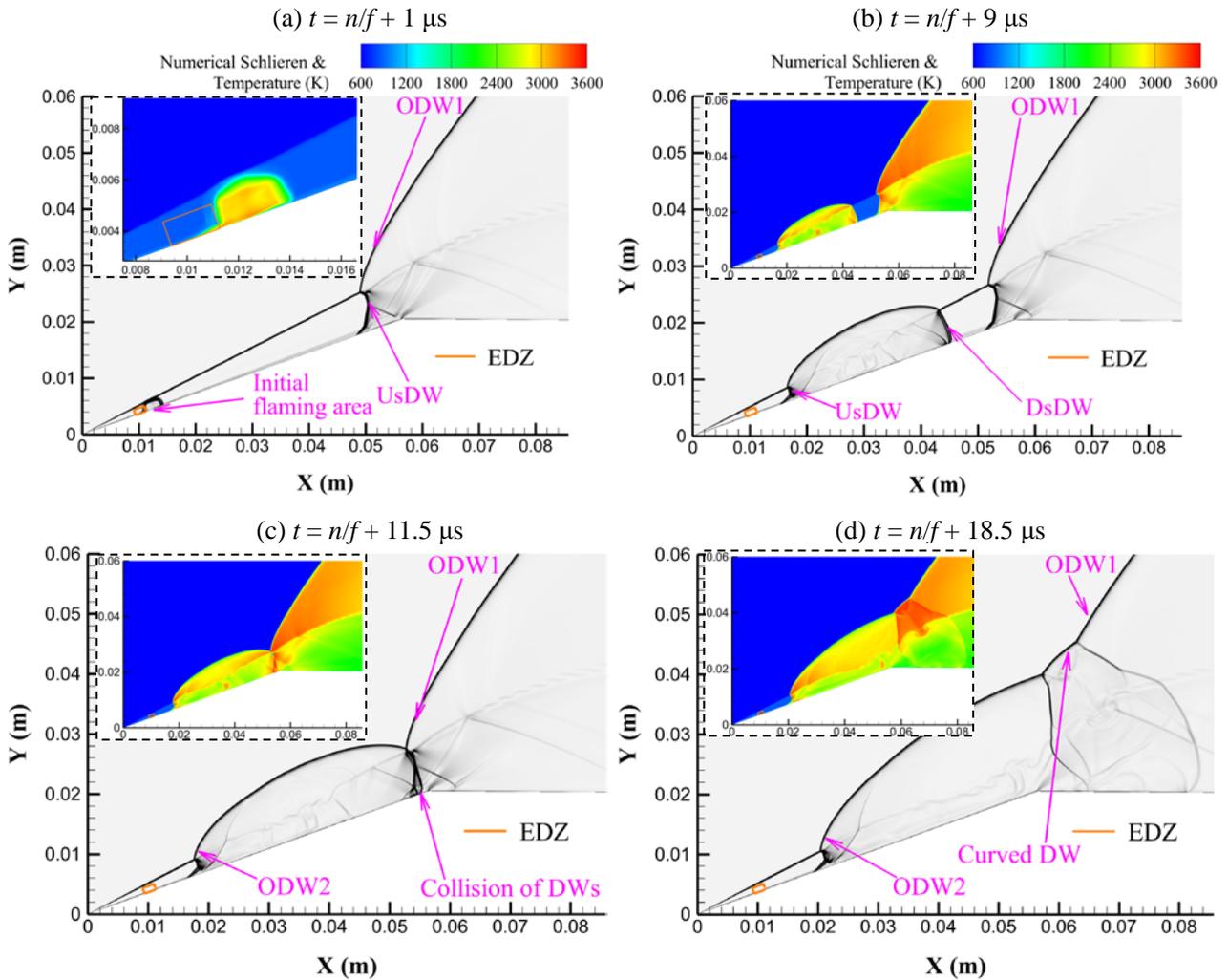

Fig. 20 Evolution of flow field after $(n+1)^{th}$ pulse in the direct explosion-spot mode under multi-pulse energy deposition with $E_{sp}$ = 50 mJ/cm and $f$ = 10 kHz. The gray contours are the numerical schlieren, while the color contours overlaid illustrate the temperature distributions.

Figure 21 compares the transient locations of the leading detonation fronts at different time instants within one repletion cycle under the direct explosion-spot mode with $E_{sp}$ = 50 mJ/cm. At $f$ = 10 kHz, which is close to the corresponding $f_{cri,DW}$, the ODW surface is basically continuous (see Fig. 21a), again demonstrating the idea described in Fig. 18 by taking the collision of DWs before the expansion corner as the sufficient condition for sustained on-wedge initiation of ODW in the direct explosion-spot mode. At a higher frequency of $f$ = 50 kHz, which is about five times $f_{cri,DW}$ (see Fig. 21b), the ODW surface is not only continuous but also nearly straight with small-amplitude oscillations. Because of this feature, the direct explosion-spot mode in multi-pulse energy deposition is suitable for practical applications.

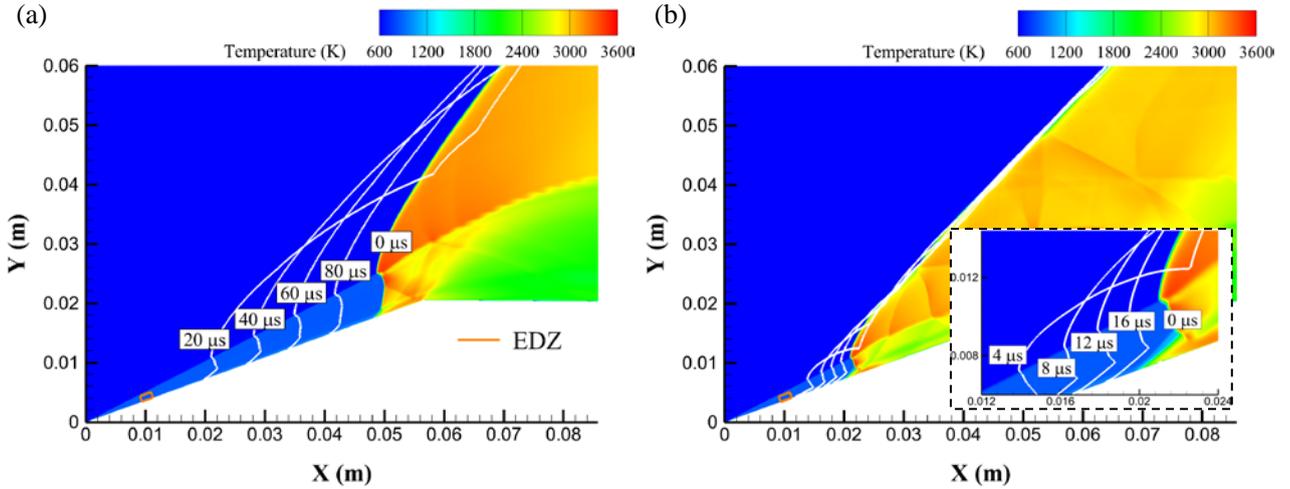

Fig. 21 Comparison of the locations of the leading detonation fronts at different instants within the $(n+1)^{th}$ cycle under multi-pulse energy deposition with $E_{sp}$ = 50 mJ/cm: (a) $f$ = 10 kHz and (b) $f$ = 50 kHz. The background temperature contours are given at $t = n/f + 0$ μs.

*3.3.2 Delayed explosion-spot mode by multi-pulse energy deposition*

As for the delayed explosion-spot mode with a pulse repetition frequency greater than the $f_{cri,DW}$ for on-wedge collision of the DWs, the spatiotemporal evolution of the detonation front is similar to those in direct explosion-spot mode as illustrated in Fig. 20, which is continuous with oscillations. However, an additional explosion spot is generated in the delayed explosion-spot mode when the RFs of two consecutive pulses collide with each other, forming a new DW and advancing the interaction of DWs in this mode. Taking $E_{sp}$ = 25 mJ/cm and $f$ = 50 kHz (approximately twice of the corresponding $f_{cri,DW}$ = 23.3 kHz as in Fig. 19) as an example, the DsRF generated by the $(n+1)^{th}$ pulse (i.e., DsRF2) catches up with the UsRF generated by the $n^{th}$ pulse (i.e., UsRF1) at 11.5 μs after the deposition of the $(n+1)^{th}$ energy pulse, as shown in Fig. 22a. At this instant, the SOSW from the $n^{th}$ pulse (i.e., SOSW1 in the figure) acts on the collision area of these two RFs, increasing local pressure and temperature significantly, accelerating local combustion, and ultimately generating a local explosion spot. Similar to the flow field evolution with the corresponding single-pulse energy deposition shown in Fig. 13, this explosion spot initiates a local DW that propagates downstream.

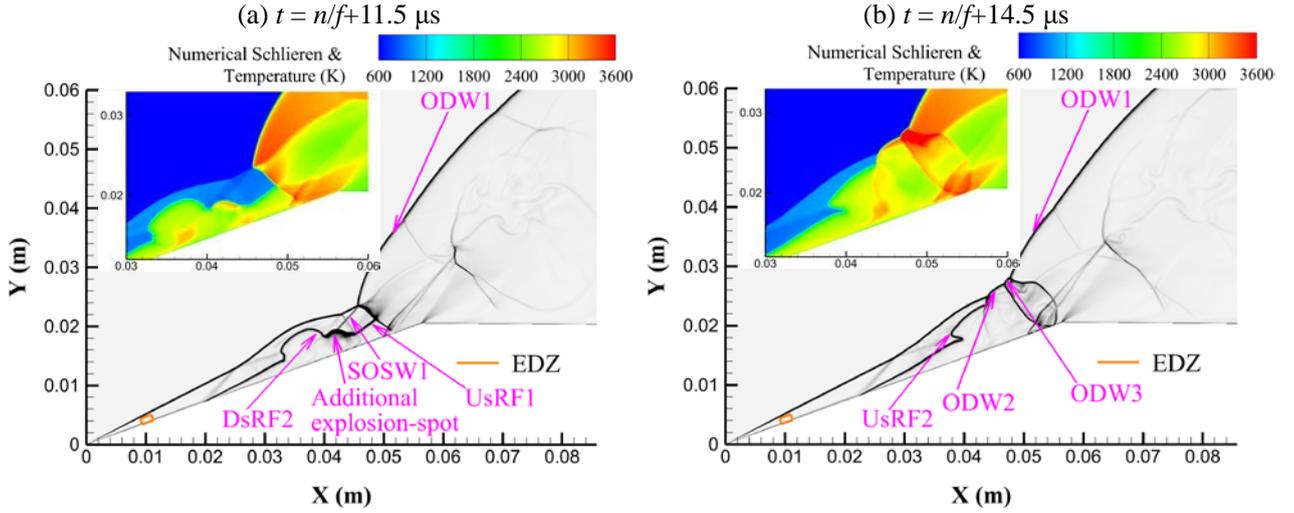

Fig. 22 Generation and evolution of the additional explosion-spot by collision of RFs between two adjacent pulses in delayed explosion-spot mode with $E_{sp}$ = 25 mJ/cm and $f$ = 50 kHz. The gray contours represent the numerical schlieren, while the color contours represent the locally enlarged temperature distributions.

As compared to Fig. 13d, the generation of this explosion spot is earlier than the generation of the original upstream explosion spot in single-pulse energy deposition. Hence, the continuous on-wedge ODW is sustained by the collision of this local DW with the ODW of the last energy pulse (see Fig. 22b), rather than the collision of DsDW and UsDW required in the direct explosion-spot mode (see Fig. 20c). In other words, only the collision of RFs is required in this mode, and the minimum repetition frequency is determined by $f_{cri,RF}$ (= 19.6 kHz at $E_{sp}$ = 25 mJ/cm), which is smaller than the corresponding $f_{cri,DW}$ (= 23.3 kHz).

The observed feature of the generation of additional explosion spots by collision of RFs can extend the energy range of the delayed explosion-spot initiation mode in multi-pulse energy deposition down to the initiation failure (but combustion happens) regime in single-pulse energy deposition. Based on the results given in Sections 3.2.1 and 3.2.2, the new energy range is 15 mJ/cm ≤ $E_{sp}$ < 27 mJ/cm. To validate this approach, multi-pulse energy deposition at $E_{sp}$ = 17 mJ/cm and $f$ = 50 kHz (approximately twice the corresponding $f_{cri,RF}$ = 25.6 kHz as in Fig. 19) is taken as an example, and the transient evolution of the flow field is shown in Fig. 23.

At the early stage after energy deposition of the $(n+1)^{th}$ pulse (see Fig. 23a), only a high-temperature zone without any combustion is created near the wall at this $E_{sp}$. At this instant, a long RF from the $n^{th}$ pulse is located downstream, following which are the UsDW and the initiated ODW. At $t = (n+1)/f+10$ μs, as shown in Fig. 23b, decoupled combustion is induced, and the RF from the $(n+1)^{th}$ pulse quickly propagates downstream. It collides with the RF from the $n^{th}$ pulse at $t = (n+1)/f+13.5$ μs (see Fig. 23c), and a local explosion spot is generated in the collision area of the RFs of these two consecutive pulses. As a result, a curved detonation wave is initiated from this explosion spot and quickly propagates downstream (see Fig. 23d). At $t = (n+1)/f+17.5$ μs (see Fig. 23e), this curved DW collides with the previous ODW generated by the $n^{th}$ pulse. Ultimately, additional ODW segments are generated, which sustains the continuity of the oblique detonation fronts, as shown in Fig. 23f.

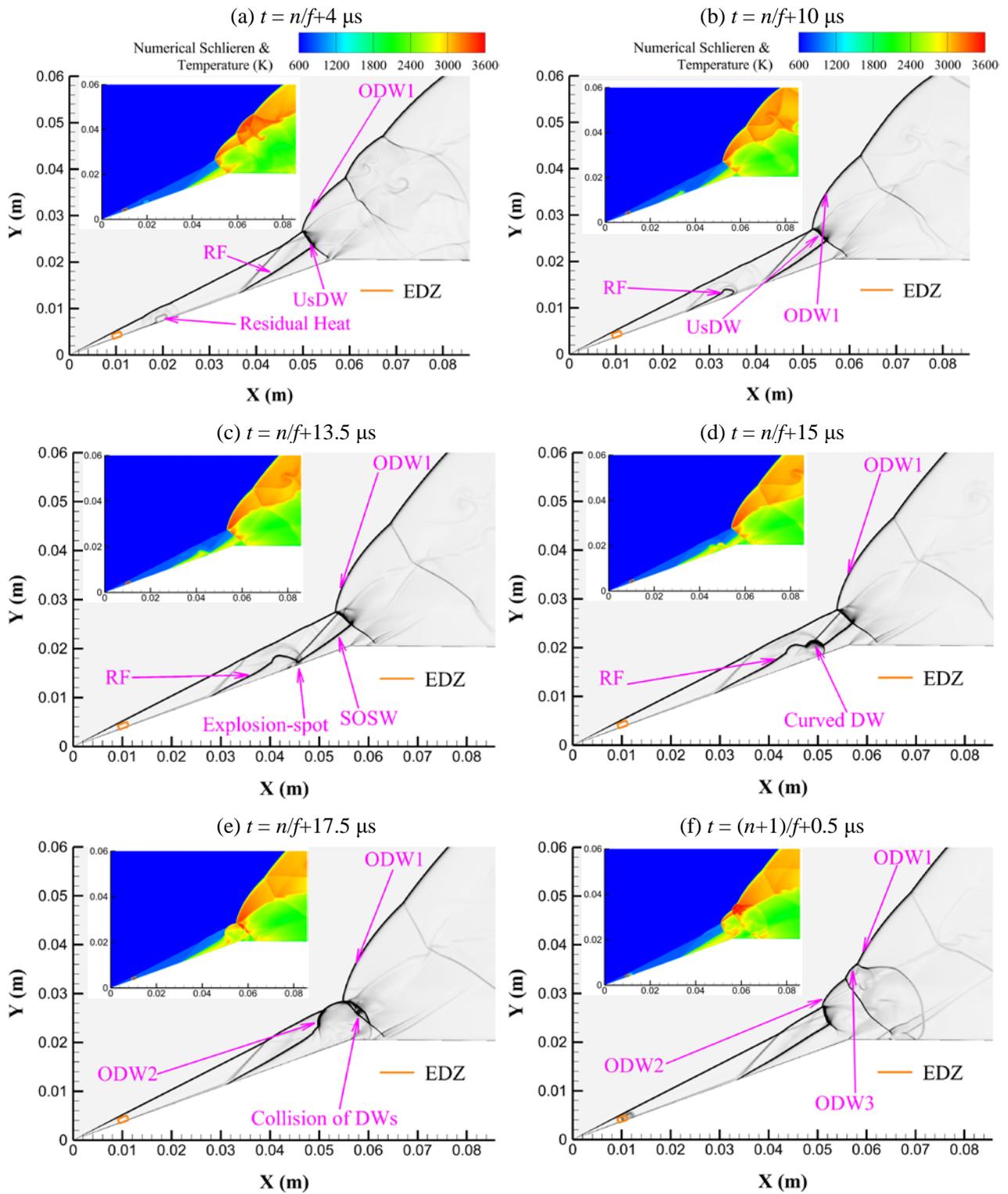

Fig. 23 Early on-wedge initiation by the collision of RFs between two adjacent pulses with $E_{sp}$ = 17 mJ/cm and $f$ = 50 kHz.

*3.3.3 Averaged input power in multi-pulse energy deposition*

In multi-pulse energy deposition, the averaged input energy can be easily calculated by the product of pulse energy and pulse repetition frequency, i.e., $\bar{P}_{in} = E_{sp} \cdot f$. For the comparison of the initiation characteristics of ODWs assisted by the continuous and pulsatile forms of local energy depositions in terms of required input energy, it should be done under equivalent initiation length. In continuous energy deposition, the location of MP is used as the initiation length, i.e., $L_{mp}$, as defined in Section 3.1.2. As for pulsatile energy deposition, the location of MP evolves in time. In this case, the mean value of $L_{mp}$ within one pulse, i.e., $\bar{L}_{mp}$, is employed as the equivalent initiation length.

Figure 24 compares the variations of the equivalent initiation length with the average input power between the continuous and pulsatile forms of energy deposition. Generally, the required average input power of the pulsatile energy deposition is smaller than that of the continuous energy deposition by more than one order of magnitude for the same ODW initiation length. For example, for an equivalent initiation length of 0.06 m, the required average input power of the pulsatile energy deposition ($\bar{P}_{in}$ = 0.85 kW/cm with $E_{sp}$ = 17 mJ/cm and $f$ = 50 kHz) is approximately 6.5% of that of the continuous one ($P_{in}$ = 13 kW/cm). For an equivalent initiation length of 0.036 m, the required $\bar{P}_{in}$ by using pulsatile energy deposition with $E_{sp}$ = 50 mJ/cm and $f$ = 10 kHz is 0.5 kW/cm, which is only 2.3% of that by using continuous energy deposition (22 kW/cm). The above results demonstrate that the pulsatile form of energy deposition is more practical than the continuous one in terms of energy consumption.

Furthermore, the three cases of on-wedge initiation of ODW assisted by pulsatile energy deposition under the direct explosion-spot mode have smaller equivalent initiation lengths than those of the two cases under the delayed explosion-spot mode, but the required average input powers are in a similar order of magnitude. For example, by employing $E_{sp}$ = 50 mJ/cm and $f$ = 20 kHz, the equivalent initiation length is reduced by approximately 40% as compared to that using $E_{sp}$ = 25 mJ/cm and $f$ = 50 kHz. This implies that for the same average input power, pulsatile energy deposition with a low pulse repetition frequency but a high pulse energy to implement the direct mode is more favorable in practical applications in terms of ODW initiation.

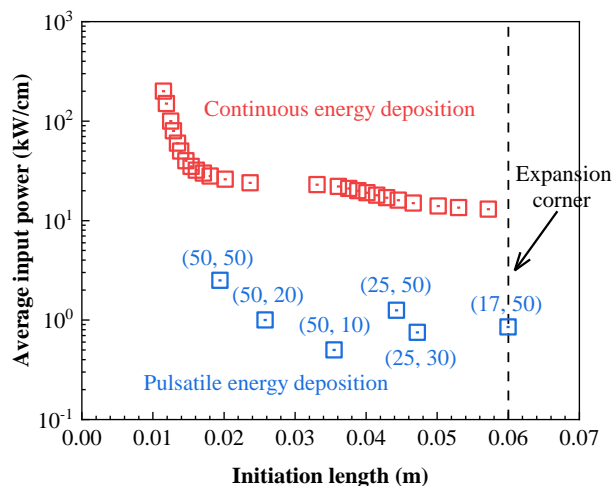

Fig. 24 Comparison of equivalent initiation length between pulsatile energy deposition and continuous energy deposition at different average input power. The data marked near the points in pulsatile energy deposition is given in ($E_{sp}$, $f$) with units in (mJ/cm, kHz).

# 4. Concluding remarks

Failures in ODW initiation could happen when the ODEs operate under extreme flight conditions of low Mach numbers and/or high altitudes, calling for the utilization of extra initiation assistance measures in the combustors. In this study, the thermal effects of plasma-based techniques for ODW initiation assistance using continuous and pulsatile types of plasma sources are numerically studied by introducing local energy deposition into the flow field over a finite wedge. Results show that both continuous and pulsatile local energy depositions can initiate ODWs effectively, resulting in sustainable detonation over the finite wedge. Moreover, the detailed initiation mechanisms of ODWs assisted by continuous and pulsatile local energy depositions are revealed by examining the steady-state flow structures with different deposition powers and the transient evolutions of key flow structures with different pulse energies, respectively.

Under continuous local energy deposition, three main modes occur in sequence as the deposition power $P_{in}$ increases, namely, the initiation-failed mode, the delayed initiation mode, and the direct initiation mode. In the delayed initiation mode, combustion is not induced by the energy deposition immediately, but happens downstream of the EDZ. Thereafter, ODW initiation is achieved through combustion acceleration. Based on the different variations of the RF and MP, the delayed mode can be further divided into sub-mode A and sub-mode B, as schematically summarized in Fig. 25. As for sub-mode A, the combustion acceleration is gradual and caused by multiple shock compressions of OSW, SOSW1, and SOSW2. In this case, the RF is long, and the locations of the RF and the MP move upstream synchronously as $P_{in}$ increases. As for sub-mode B, a much faster acceleration of combustion is induced by the higher energy deposition as well as the shock compressions of OSW and SOSW1. As a result, the RF transits into a SODW (rather than the SOSW2) immediately as it goes across the SL, leading to a sudden advancement of the MP and the transition of sub-mode A to sub-mode B by increasing $P_{in}$. In direct initiation mode (see Fig. 25), combustion and detonation initiation are directly triggered by the massive amount of energy deposited into the EDZ, and compression from any SOSWs, as in the delayed mode, is not required. This feature makes ODW initiation happen with a very short distance or even take place within the EDZ, and hence, the initiation location of ODW in this mode is controllable by the placement of EDZ based on practical requirements in applications, which is favorable.

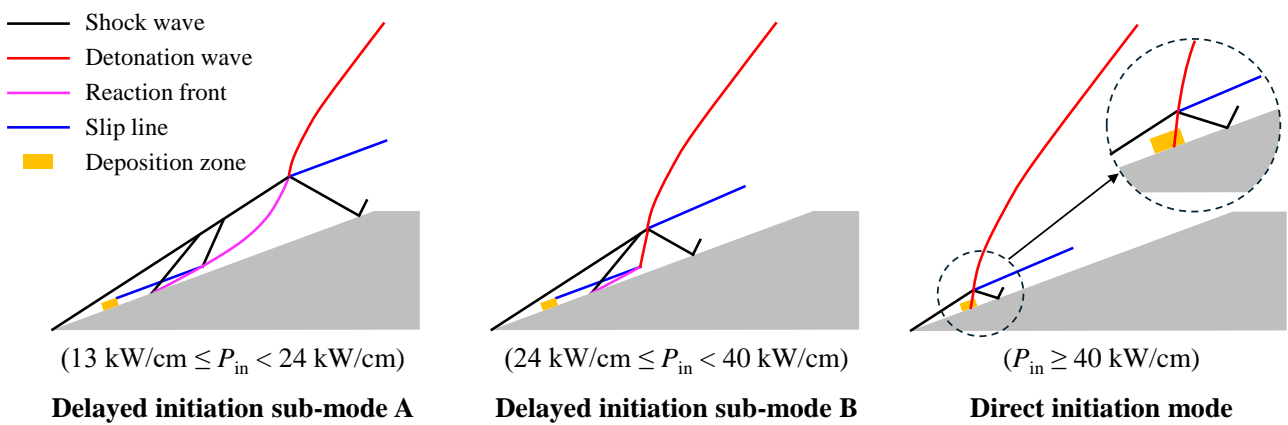

Fig. 25 Different ODW initiation modes assisted by continuous energy deposition.

Under pulsatile local energy deposition, the ODW is generated from the formation of a local explosion spot. As the pulse energy $E_{sp}$ increases, similar delayed and direct modes can be observed, as schematically summarized in Fig. 26. In the delayed mode with single-pulse energy deposition, an upstream local explosion spot is formed by the interaction of the reflected EW and the RF downstream of the EDZ, as for the one with multi-pulse deposition, a downstream local explosion spot is formed through collision of RFs from two consecutive energy pulses, which happens earlier than the formation of the upstream explosion spot. Hence, the existence of the delayed mode with multi-pulse deposition is extended to lower $E_{sp}$ compared to that with single-pulse deposition. In the direct mode, the local explosion spot is formed immediately as the energy pulse is deposited into the EDZ. Through analysis of the spatiotemporal evolution of the primary wave structures under single-pulse deposition, the minimum pulse repetition frequencies required for on-wedge collision of RFs and DWs from two consecutive energy pulses, i.e., $f_{cri,RF}$ and $f_{cri,DW}$, are respectively revealed, and it is demonstrated that the sustainable on-wedge ODW initiation with the delayed mode is controlled by $f_{cri,RF}$ and direct mode by $f_{cri,DW}$ (see Fig. 26). Finally, it is found that the average input power required for sustainable on-wedge detonation by pulsatile energy deposition is smaller than that by continuous deposition by at least one order of magnitude while maintaining the same equivalent initiation length, suggesting that the pulsatile energy deposition (corresponding to pulsatile type of plasma sources) is a more efficient way for ODW initiation facilitation. The results demonstrate the effectiveness of plasma-assisted initiation techniques for ODWs, providing insights for future design of ODEs operating under wide-range flight conditions.

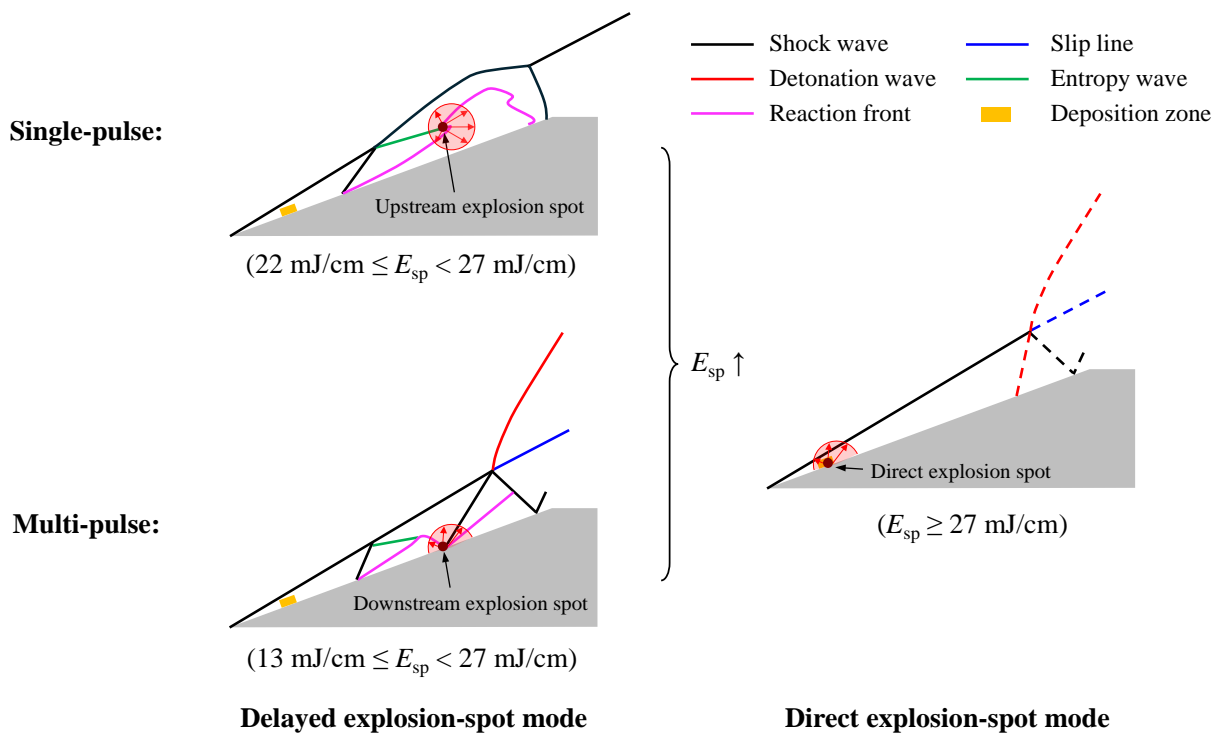

Fig. 26 Different formation mechanisms of the explosion spot in pulsatile energy deposition.

## CRediT authorship contribution statement

**Ziqi Jiang:** Writing – original draft, Methodology, Formal analysis, Data curation, Investigation, Visualization. **Zongnan Chen:** Writing – review & editing, Methodology, Software. **Lisong Shi:** Writing – review & editing, Investigation. **Zijian Zhang:** Writing – review & editing, Conceptualization, Methodology, Supervision, Resources, Funding acquisition, Software, Validation. **Jiaao Hao:** Writing – review & editing, Software, Resources, Funding acquisition. **Chih-yung Wen:** Writing – review & editing, Supervision, Resources, Funding acquisition.

## Declaration of competing interests

The authors declare that they have no known competing financial interests or personal relationships that could have appeared to influence the work reported in this paper.

## Acknowledgments

This work was supported by the National Natural Science Foundation of China (Grant Nos. 12202374 and 12472239) and the Hong Kong Research Grants Council (Grant Nos. 15204322).